\documentclass[12pt,preprint]{aastex}

\usepackage{color}
\newcommand{\blue}{\textcolor{black}}
\newcommand{\green}{\textcolor{black}}
\newcommand{\red}{\textcolor{black}}

\shorttitle{{\it RXTE} observations of Low-Mass X-Ray Binary 4U 1608--522}
\shortauthors{Takahashi, Makishima \& Sakurai}

\begin{document}


\title{{\it RXTE} Observations of \green{the} Low-Mass X-Ray Binary \\
4U 1608--522 in Upper-Banana State}

\author{Hiromitsu Takahashi}
\affil{Hiroshima Astrophysical Science Center, Hiroshima University,
1-3-1 Kagamiyama, Higashi-Hiroshima, Hiroshima 739-8526, Japan}
\email{hirotaka@hep01.hepl.hiroshima-u.ac.jp}
\and
\author{Soki Sakurai and Kazuo Makishima}
\affil{Department of Physics, University of Tokyo,
7-3-1 Hongo, Bunkyo-ku, Tokyo 113-0033, Japan}

\begin{abstract}
To investigate the physics of mass accretion onto weakly-magnetized neutron stars,
{95} archival {\it RXTE} data {sets} of an atoll source 4U 1608--522,
{acquired over 1996-2004} in so-called upper-banana state, were analyzed.
{The object meantime exhibited 3--30 keV luminosity in the range of 
$\lesssim 10^{35} - 4 \times 10^{37}$ erg s$^{-1}$, assuming a distance of 3.6 kpc.}
 {The 3--30 keV PCA  spectra, produced one from each dataset,}
were represented successfully with a combination of a soft and a hard component,
{of which the presence  was revealed in a model-independent manner
by studying spectral variations among the observations.}
The {soft component} is expressed by so-called 
multi-color disk model with a temperature of $\sim 1.8$ keV,
and is attributed to the emission from an optically-thick standard accretion disk.
The {hard component} is a blackbody emission 
{with a temperature of} $\sim 2.7$ keV,
thought to be emitted from the neutron-star surface.
As the total luminosity increases,
a continuous decrease was observed in the ratio of 
the blackbody luminosity to that of the disk component.
This property suggests that the matter flowing through the accretion disk 
gradually becomes difficult to reach the neutron-star surface, 
presumably forming outflows driven by the increased radiation pressure.
{On time scales of hours to days,
the overall source variability was found to be 
controlled by two independent variables;
the mass accretion rate, 
and  the innermost disk radius
which changes both physically and artificially.}
\end{abstract}


\section{INTRODUCTION}

It has long been known that  low-mass X-ray binaries (LMXBs),
namely close binaries involving neutron stars (NSs) without strong magnetic fields,
{emit thermal type spectra
when they are luminous ($\gtrsim 10^{36}$ erg s$^{-1}$)}.
In the 1970's, {their spectra}
were often modeled 
empirically in terms of simple thermal Bremsstrahlung, without much physical basis.
As observations made progress, it gradually became clear 
that this simple empirical modeling is in fact inadequate, 
and at least two components are needed to describe 
\green{wide-band (typically 2--30 keV) LMXB spectra.
Today, a  canonical  model (sometimes called ``Eastern model'') 
interprets the X-ray spectra of  a luminous LMXB 
as originating from two major emission regions;
the surface of the NS \green{(or often called ``boundary layer'')}
and an accretion disk around it \citep{mitsuda_z}.}

The Eastern model was \green{constructed} based on {\it Tenma} observations of 
luminous LMXBs \citep{mitsuda_z,mitsuda_1608,makishima_gx3+1},
particularly considering intensity-correlated spectral changes.
{When an LMXB becomes brighter on a time scale of $\sim$ 1000 s,
its spectrum  hardens in energies from $\sim 2$ to $\sim 10$ keV,}
but the spectral shape in the hardest range (above 10 keV) stays constant.
A difference between a pair of spectra with different intensities generally 
{takes a form of \red{a blackbody (BB) \blue{spectrum}},
of which} the intensity changes but the shape (i.e., the temperature) is approximately constant.
{This BB component has a temperature of $\sim$ 2 keV,
which} is close to the Eddington temperature, 2.0 keV, 
for a NS of $\sim$ 1.4 $M_{\odot}$ mass (with $M_{\odot}$ being the solar mass)
\blue{and 10 km radius.}
{In addition, the area of its emission region is inferred to be 
a fraction of the surface area of \blue{such a NS}.}
Therefore, the BB component has been ascribed successfully to the NS surface emission.
Indeed, every spectrum observed from these sources was reproduced 
by a sum of this BB component, and an additional soft component.
The soft component was approximated first by a softer BB of $\sim 1$ keV temperature,
but later, much better by a particular superposition of BB spectra,
called ``disk blackbody" or ``multi-color disk" (MCD) emission \citep{mitsuda_z,makishima_gx3+1},
as predicted by the optically-thick standard accretion disk model \citep{shakura_disk}.

While the Eastern model thus provides a promising ground to 
understand  the physics of mass accretion {in luminous LMXBs},
their complex spectral and intensity variations have often been described 
in a purely empirical manner using ``color-color'' diagrams and other similar methods.
As a result, such empirical classifications as ``Z sources'' 
and ``atoll sources'' have been created,
together with various ``branches'' on these empirical diagrams \citep[e.g., ][]{hasinger_z}.
These primitive descriptions are waiting to be replaced by more physically meaningful ones,
using \blue{the Eastern model}.

After the launch in 1995, the {\em Rossi X-Ray Timing  Explorer} 
\citep[{\it RXTE};][]{bradt_rxte} has been observing many 
LMXBs for a huge number of times.
With the largest effective area and the highest timing resolution ever achieved,
{\it RXTE} is {indeed} very suitable to the study of 
{spectral variations} of LMXBs.
{On time scales of milliseconds to seconds,
\citet{gilfanov_vari} and \citet{revnivtsev_vari}
carried out such studies using the {\it RXTE} data, and revealed
that the variation is carried by a rather hard  BB-like emission component
 with a temperature of $\sim$ 2--3 keV.}
These results reinforce the validity of the Eastern model,
and encourage us to attempt its extensive application to the {\it RXTE} data of LMXBs.
{We thus} hope to understand the spectral variations of LMXBs
as a whole using a physical model (based on the Eastern description).

\if
Contrary to the Eastern model, the Western model assumes that Thomson opacity dominates free-free opacity in the accretion disk and the emission is significantly modified by electron scattering, although the emission from the NS surface are hardly modified and observed as simple BB emission.
\citet{white_west} for the first time applied this model to actual X-ray data, and argued that it can reproduce the observed LMXB spectra best among many models they attempted, including the Eastern model without the consideration of the Comptonization hard tail.
As a variant to this model, there have been repeated suggestions that the accretion disk is surrounded by a cloud of hot electrons, or ``accretion disk corona'', 
which may add to the Comptonization.
\citet{church_west} analyzed dip sources, and also argued that the Western model can reproduce their spectra in a unified way at various dipping levels, and hence with different shapes.
\fi

As the first of a series of {our planned} publications,
the present paper deals with so-called upper-banana state {of 4U 1608--522,
which is one of  the most frequently observed  atoll sources}  by {\it RXTE}.
{Section 2 describes  414  {\it RXTE} data sets of  this LMXB,
of which 95 are selected for use in the present paper.
In \S~3.1, we select four representative spectra out of the 95 data sets,
and  analyze  their intensity-correlated spectral changes
to reconstruct the Eastern model.
The derived model parameter values are examined in \S~3.2,
followed by \S~3.3 where the model  is applied  to all the 95 energy spectra
and  relations among the obtained physical parameters are studied.
As a reconfirmation of our approach,
we calculate in \S~3.4 the effective degrees of freedom
involved in the variability,
of which the results are physically interpreted in \S~3.5.
As discussed in \S~4,
the results} give a full support to the Eastern model,
and lead to a finding of mass outflows as the source gets luminous.

\section{OBSERVATIONS AND DATA REDUCTION}

{Our target object 4U 1608--522 is an LMXB 
with a recurrent transient characteristic,  
exhibiting occasional outbursts.
As  the X-ray intensity evolves through these outbursts,
the source  is known} to take all three spectral states of atoll sources \citep{hasinger_z};
namely island, lower-banana, and upper-banana states,
in the increasing order of the source intensity \citep{muno_z_atoll,gierlinski_z_atoll}.
In the island and lower-banana states,
a hard tail appears in the spectra presumably due to Comptonization by hot electrons,
of which the temperature is of the order of several tens keV \citep{gierlinski_1608-522}.
In the present paper, we analyze only the data in the upper-banana state,
where the spectrum takes the thermal-shape characteristic of luminous LMXBs.
The original paper by \citet{mitsuda_z} also utilized 4U 1608--522,
when the source was presumably in the upper-banana state.
The source distance is assumed to be 3.6 kpc \citep{nakamura_hard}.

{Although}  {\it RXTE} has observed many outbursts from 4U 1608--522,
we limit {the present study to those data sets which were obtained}
from the launch of {\it RXTE} (December 1995) to the end of August 2004,
when the brightest and the second brightest outbursts {were recorded.}
During the period,
414 pointing observations of 4U 1608--522 were conducted 
with the Proportional Counter Array 
\citep[PCA;][]{jahoda_pca}.
Although some of the five Proportional Counter Units (PCUs) 
of the PCA were turned off in some observations,
PCU2 alone was always operational.
In order to avoid systematic {differences} among different PCUs,
we utilized the PCU2 data only,
because the source in the upper-banana state is very bright
(typically $\sim 1000$ counts s$^{-1}$ per PCU)
and hence statistical errors are usually negligible.
We employed ``Standard-2'' data, of which the time resolution is 16 s,
since they are most accurately calibrated and suitable for spectral analysis.
The data were filtered in the standard manner for bright sources,
and were corrected for dead times in the standard way.
The PCA background was estimated for each dataset by PCABACKEST version 3.0.
To remove Type I bursts, 
we excluded those time regions
when the background-subtracted source count rates (per 16 s) 
are outside 50\%--200\% of the average over each observation.

\begin{figure}[htbp]
\begin{center}
\includegraphics[width={0.55\textwidth}]
{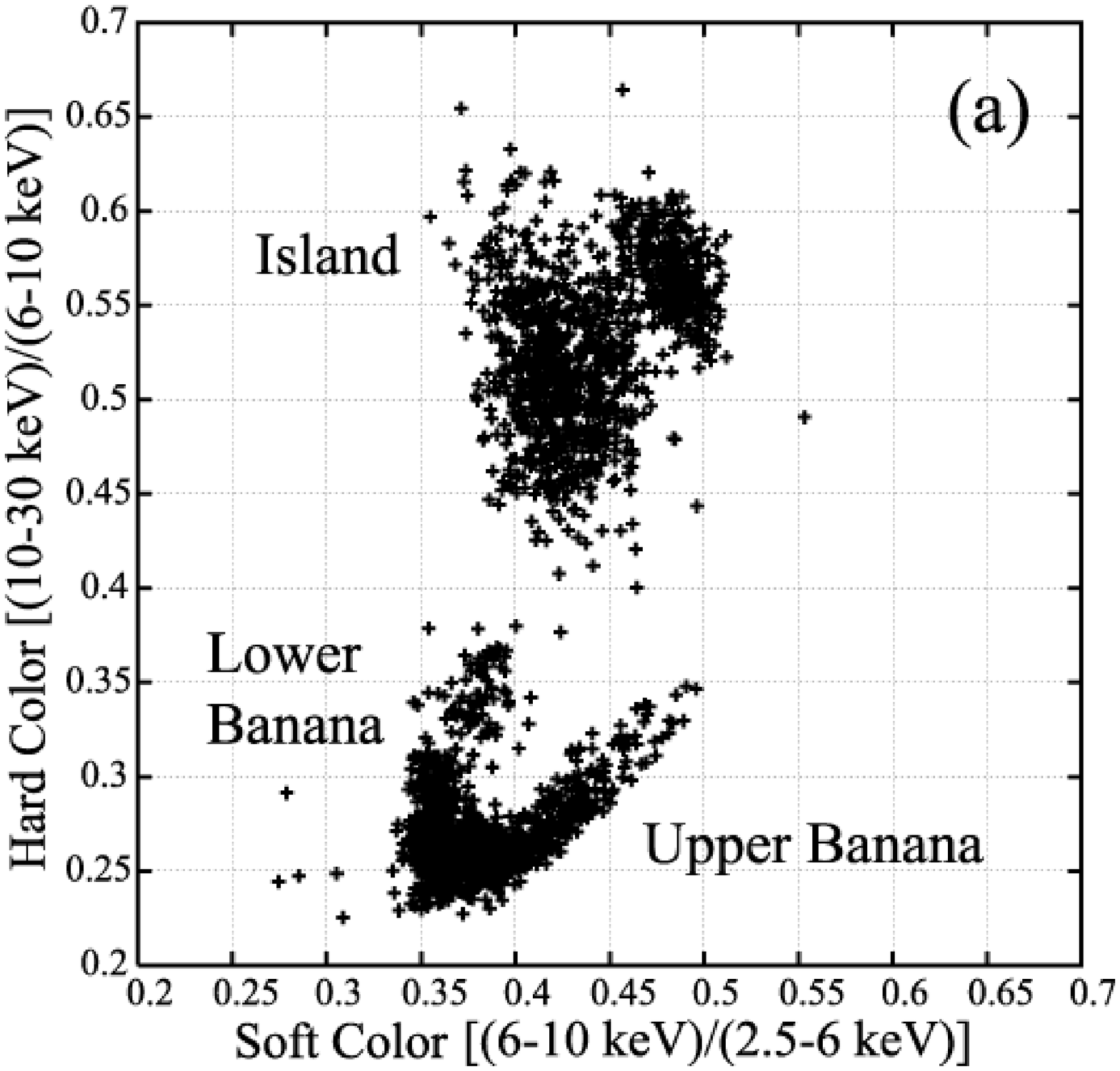}
\includegraphics[width={0.55\textwidth}]
{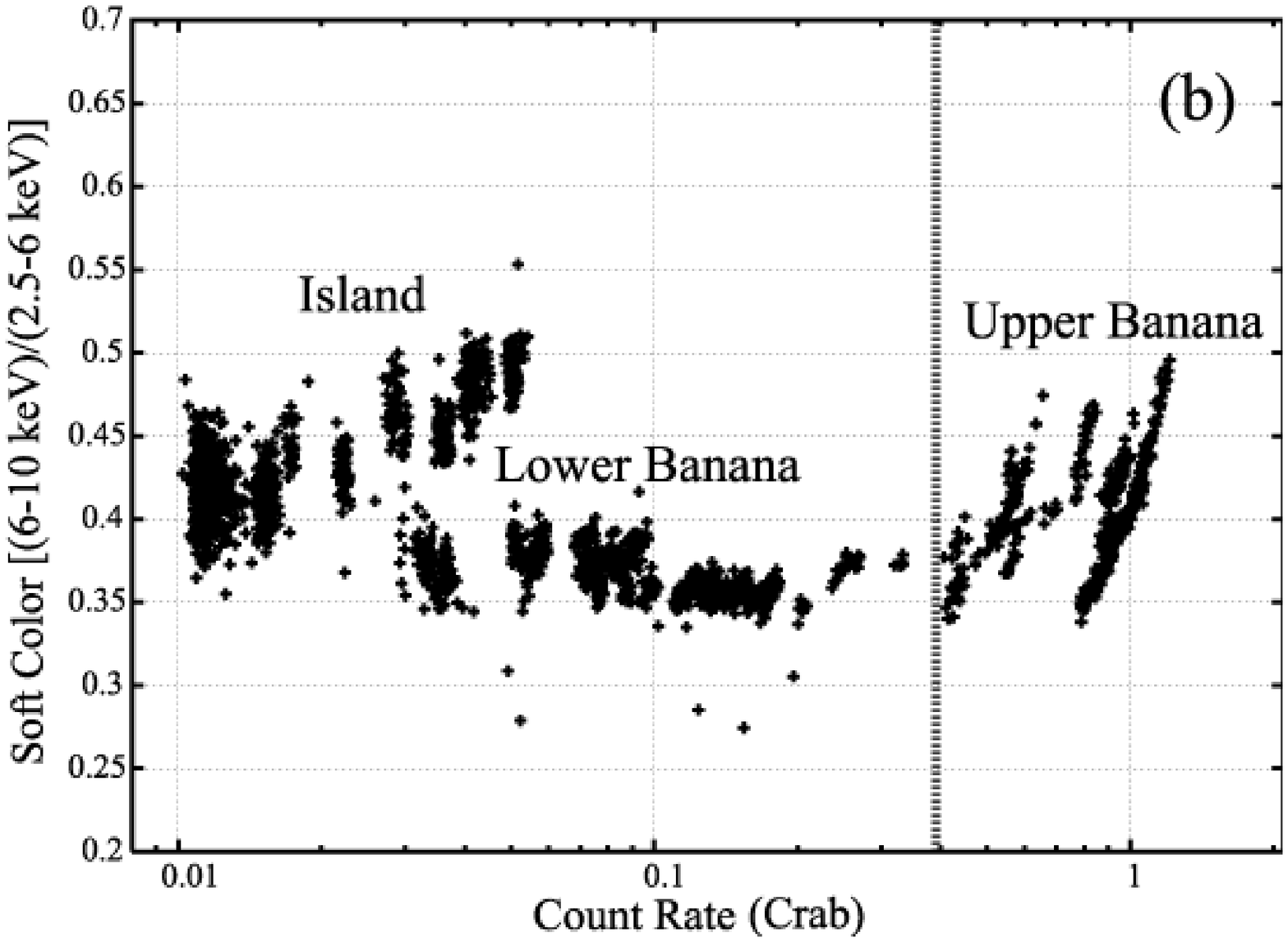}
\includegraphics[width={0.55\textwidth}]
{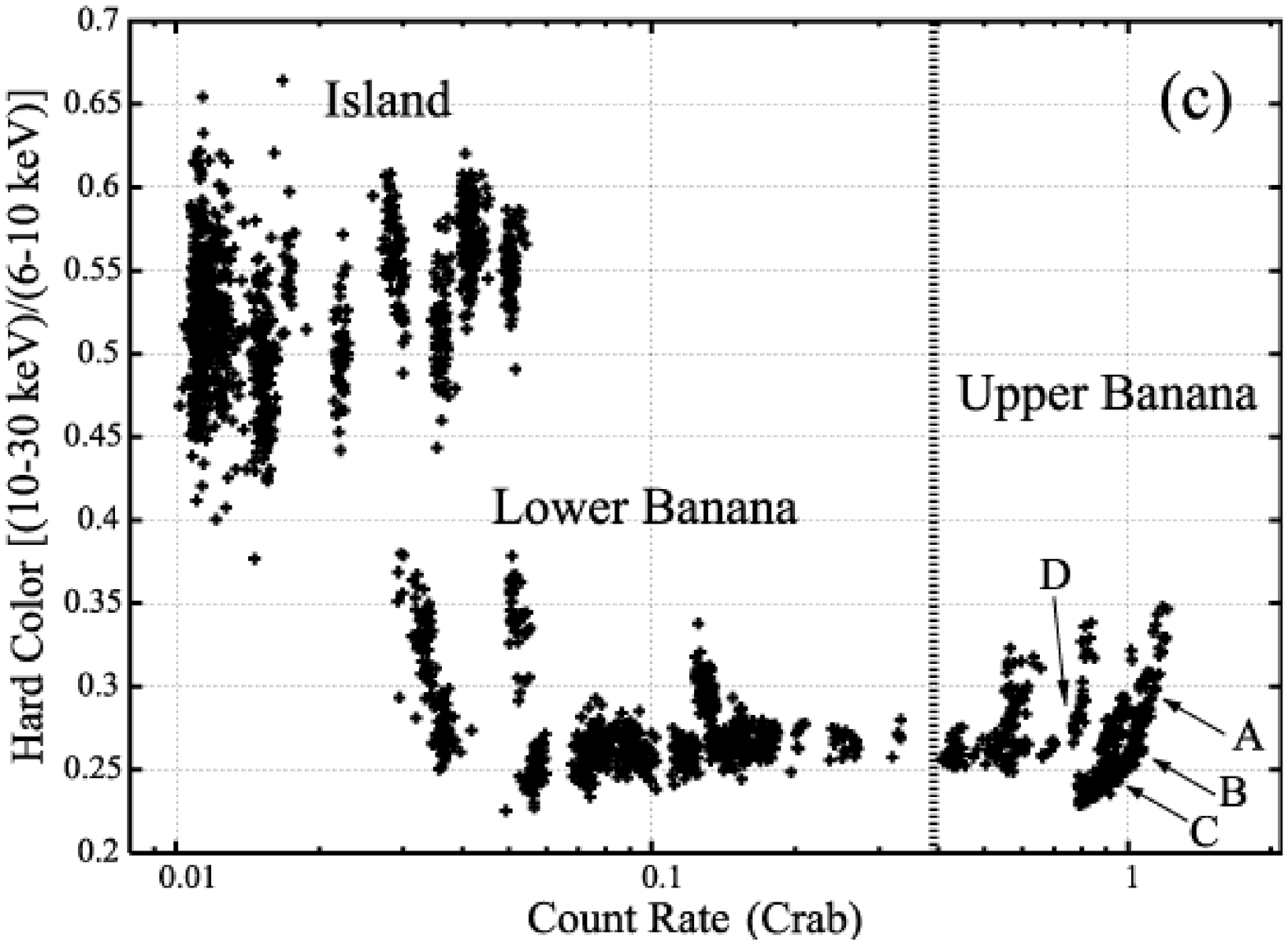}
\caption
{The CCD (panel a), soft-HID (panel b), and hard-HID (panel c) of 4U 1608--522 in epoch 3, 
with the time bin of 128 s.
The utilized energy bands are described in the text.
Four arrows indicate the datasets analyzed in \S~3.1 (Table~\ref{tab:list}).}
\label{fig:ccd_1608-522}
\end{center}
\end{figure}

After the subtraction of the modeled background,
we calculated color-color diagrams (CCDs) and hardness-intensity diagrams (HIDs) 
of 4U 1608--522 using data points averaged over 128 s.
In this paper,
the soft and hard colors refer to 6--10 keV vs. 2.5--6 keV source count ratios
and those in 10--30 keV vs. 6--10 keV, respectively.
The intensity used for HIDs refers to an energy band of 2.5--30 keV
in the unit of the Crab Nebula count rate.
As described in \citet{jahoda_pca2},
the overall PCA operation history is divided into 5 epochs
according to differences in the high-voltage settings.
As a result, CCDs and HIDs need to be produced separately for different epochs.
Figure~\ref{fig:ccd_1608-522} representatively shows the obtained CCD and HIDs 
of 72 observations in the entire epoch 3.
Here, we excluded faint data points 
of which the intensities are less than 10 counts s$^{-1}$ PCU2$^{-1}$
($\lesssim 3 \times 10^{35}$ erg s$^{-1}$ for 3.6 kpc).

{On} the CCD of Figure~\ref{fig:ccd_1608-522},
the count rate increases from top through lower left to lower right,
with each part corresponding to the island, lower-banana, and upper-banana states, respectively.
On the HIDs, {the data points form  many vertical  line-like features, or stripes.}
The source traces these {stripes} on a time scale of days,
and moves horizontally on much longer time scales.
The former causes the source motion {within a single} state on the CCD,
while the latter leads to transitions among the three states.
The hard color is higher in the island state,
and decreases in the banana states.
The present paper focuses on the upper-banana state,
which is defined here as those datasets with the 3--30 keV source intensity exceeding 0.4 Crab.
We applied the same criteria to the data from the other four epochs.
When summed over the 5 epochs, the selected datasets {reach} 95 in number,
with their 3--30 keV luminosities spanning a range of (1--4) $\times 10^{37}$ erg s$^{-1}$.
Some of these datasets were already analyzed with the Eastern model 
by \citet{gierlinski_1608-522}, \citet{gilfanov_vari} and \citet{lin_aql1608}.

For each dataset in the upper-banana state,
we accumulated the Standard-2 data into a single 3--30 keV spectrum,
with a typical exposure of several ks.
Then, the modeled background mentioned above {was subtracted.}
However, this method is known to causes slight (several percents) 
over-subtraction of the background
\footnote{http://lheawww.gsfc.nasa.gov/users/craigm/pca-bkg-tdrift/}.
This effect was compensated for by rescaling the background spectrum up to 10\%,
so that its count rate matches that of the on-source data 
in the hardest 60--100 keV energy range,
where the signal count is considered negligible.
Since the source count was detected significantly up to 30 keV,
we hereafter analyze the total 95 energy spectra in the energy range of 3--30 keV.
The PCU2 response matrix was created for each observation using PCARSP version 10.1.
In order to take into account calibration uncertainties,
we add 1\% and 7\% systematic errors to each energy bin of the source and background spectra, respectively.

\section{{DATA  ANALYSIS AND RESULTS}}

\subsection{Comparison of Energy Spectra}
\label{subsec:spec}

\begin{figure}[htbp]
\begin{center}
\includegraphics[width={0.55\textwidth}]
{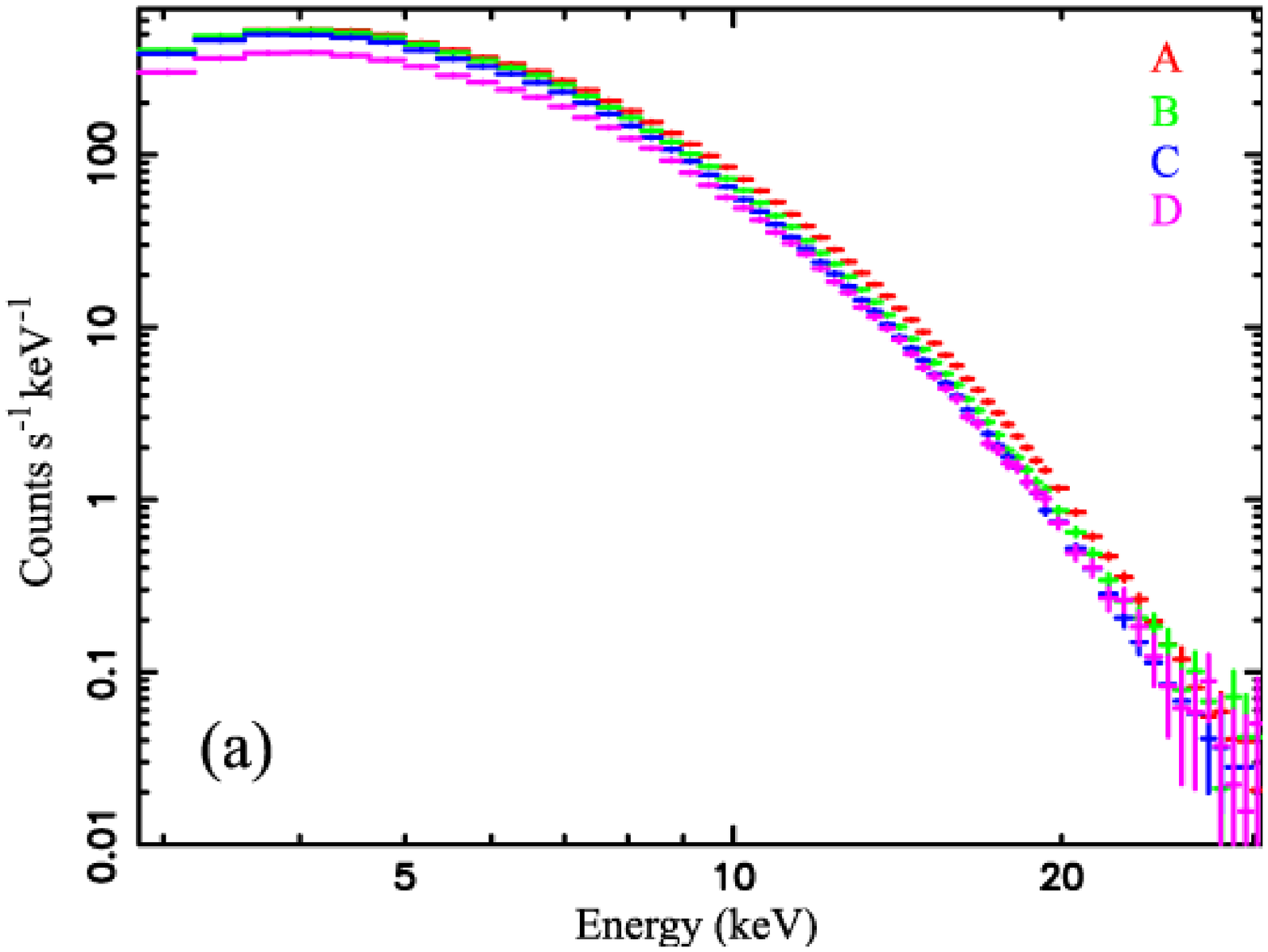}
\includegraphics[width={0.55\textwidth}]
{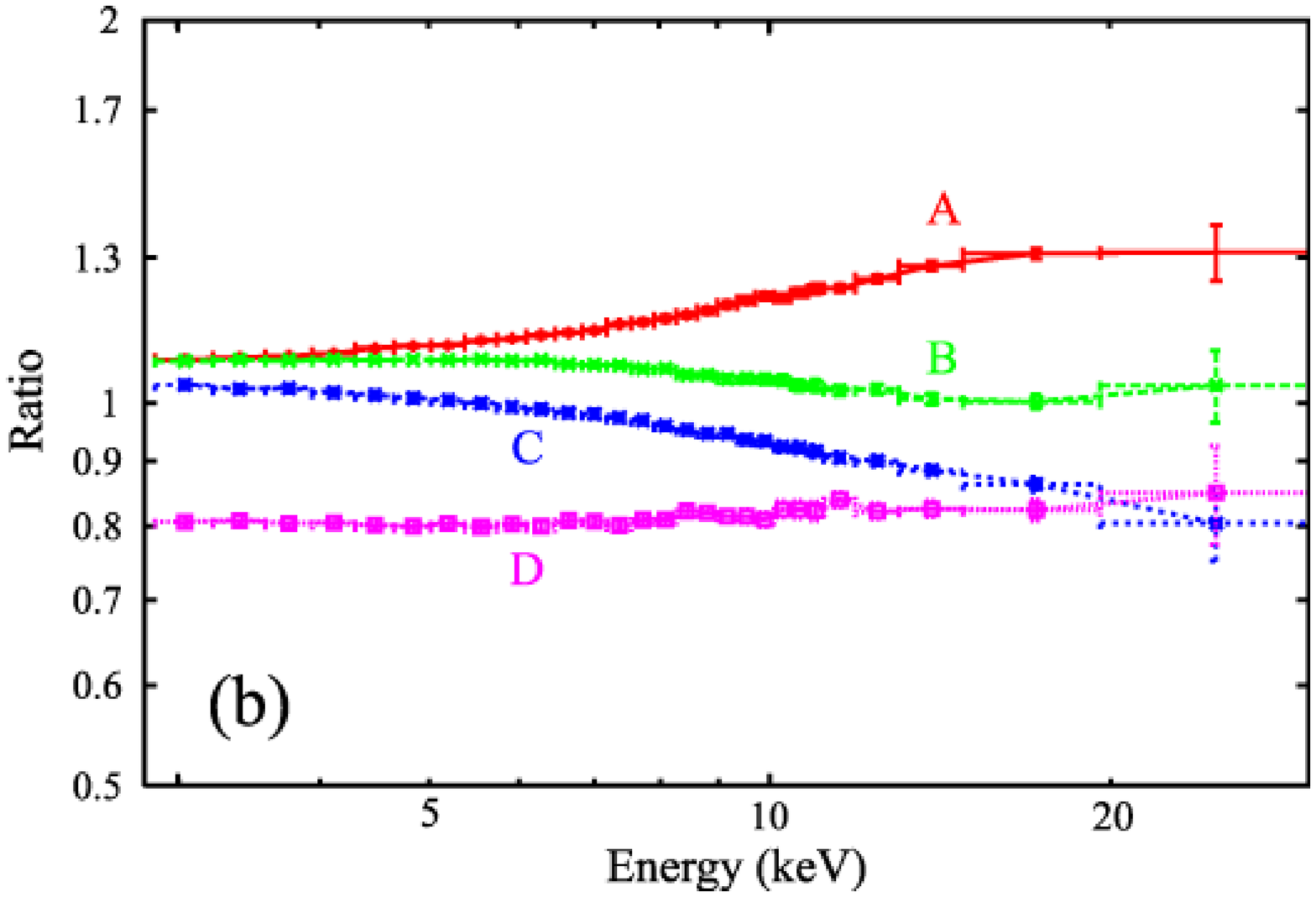}
\caption
{Four energy spectra (panel a) shown in Table~\ref{tab:list} and Figure~\ref{fig:ccd_1608-522} (c), 
and their intensity ratios to their average (panel b).
The energy spectra are shown without removing the detector response.
}
\label{fig:spec_1608_upper1}
\end{center}
\end{figure}

In comparison with the previous work by \citet{mitsuda_z},
we first examine relatively slow ($\gtrsim 10^{3}$ s) spectral variations
along the upper-banana state of 4U 1608--522.
For this purpose, we selected four representative spectra,
{hereafter Spec A through Spec D,}
obtained at the most luminous (several times $10^{37}$ erg s$^{-1}$) 
end (i.e., the right side) of the HIDs.
Three of them (Spec A, Spec B and Spec C) come from the same vertical line,
while the other is located on the different line.
Their observation IDs, average intensities, 
and soft/hard colors are summarized in Table~\ref{tab:list},
{while} their locations on the hard-HID 
are indicated in Figure~\ref{fig:ccd_1608-522}.
These spectra and their ratios to their average spectrum 
are shown in Figure~\ref{fig:spec_1608_upper1}.

\begin{figure}[htbp]
\begin{center}
\includegraphics[width={0.55\textwidth}]
{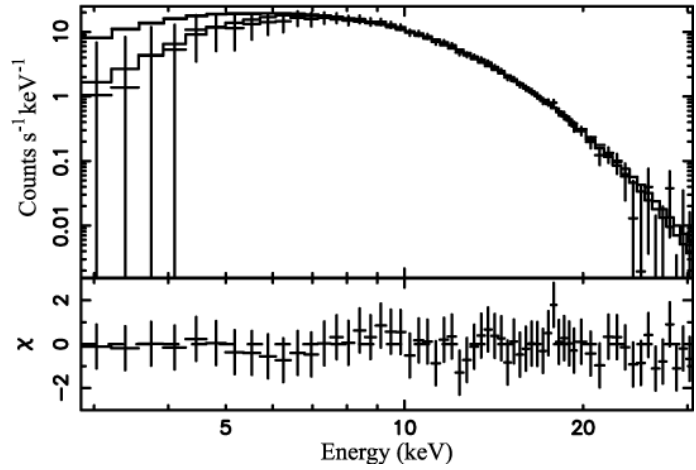}
\caption
{The difference spectrum (Spec 1) between the brightest (Spec A) 
and the second brightest (Spec B) spectra in Figure~\ref{fig:spec_1608_upper1}.
It is successfully represented by an absorbed BB model 
with a temperature $kT_{\rm BB} \sim$ 2.5 keV (lower histogram).
The fit residuals are shown in the bottom panels.
If the absorption is fixed at $0.8 \times 10^{22}$ cm$^{-2}$,
the model over-predicts the data below 7 keV (upper histogram).
}
\label{fig:spec_1608_upper2}
\end{center}
\end{figure}

In Figure~\ref{fig:spec_1608_upper1}, 
{ 
the brightest (Spec A) and the second brightest (Spec \red{B}) spectra of the four
are selected from the same vertical line in the HID.}
By comparing them,
it is clear that the intensity in the hard energy band changes significantly,
while that in the soft band is kept constant.
Moreover, in energies above $\sim$ 15 keV, 
their ratios saturates as a function of energy.
As originally proposed by \citet{mitsuda_z}, 
these results can be explained  
{when the spectra consist of a variable hard component 
that dominates in $\gtrsim 15$ keV energy band
and a stable soft component,}
and the normalization (but not the shape) of the 
{former} changes between the two spectra.
Indeed, the difference spectrum between them, 
shown in Figure~\ref{fig:spec_1608_upper2} as Spec 1, 
is represented successfully by a BB model 
of which the temperature is $kT_{\rm BB} \sim$ 2.5 keV.
{The obtained best-fit parameters are summarized in Table~\ref{tab:spec_1608_upper}.
Although the derived absorption column density,
{$\sim 3 \times 10^{23}$ cm$^{-2}$,}
is significantly higher than that reported previously 
\citep[$0.8 \times 10^{22}$ cm$^{-2}$;][]{grindlay_nh,rutledge_nh},
the difference may be attributed to} slight changes in the assumed stable soft component 
\green{(as assessed later in this subsection),}
and/or those in $kT_{\rm BB}$.
 {These results reconfirm 
the achievements with  {\it Tenma} \citep{mitsuda_z}
and {\it Ginga} \citep{makishima_gx3+1}.}

From the above results, we consider 
that the harder part of the spectrum is carried by a ``hard component'' 
which is approximated by a $kT_{\rm BB} \sim$ 2.5 keV BB.
Its variation  
{\citep[in normalization rather than in $kT_{\rm BB}$; ][]{makishima_gx3+1} } 
causes significant changes in the hard color, 
but does not affect the intensity very much 
because the total source counts are dominated by softer photons.
This explains the formation of the vertical stripes 
in the hard HID of Figure~\ref{fig:ccd_1608-522}.
This hard component is most naturally attributed to 
optically-thick emission from the NS surface, 
because the measured temperature agrees with the local Eddington 
temperature of a 1.4 $M_{\odot}$ NS, or $\sim 2.0$ keV.

\begin{figure}[htbp]
\begin{center}
\includegraphics[width={0.55\textwidth}]
{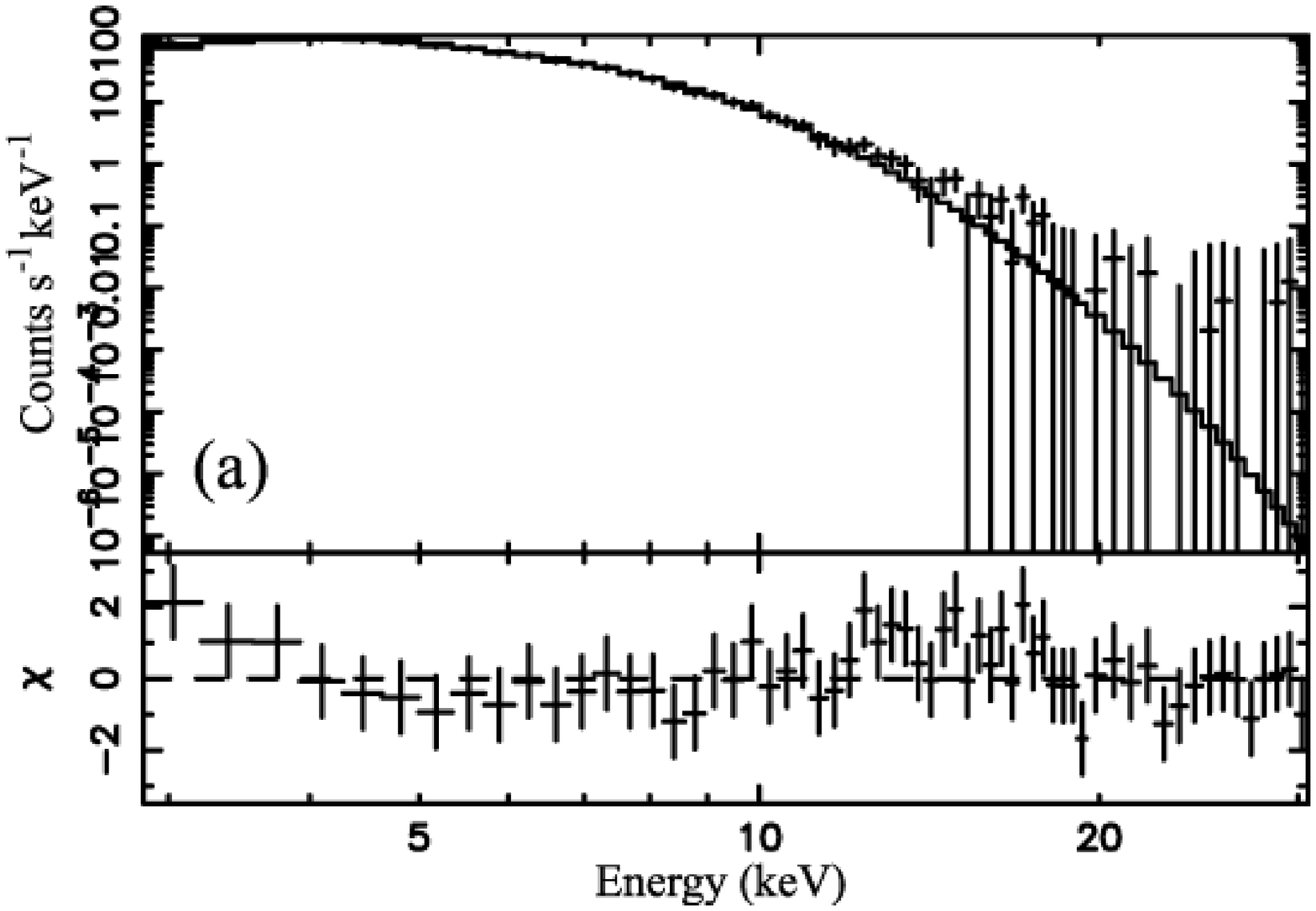}
\includegraphics[width={0.55\textwidth}]
{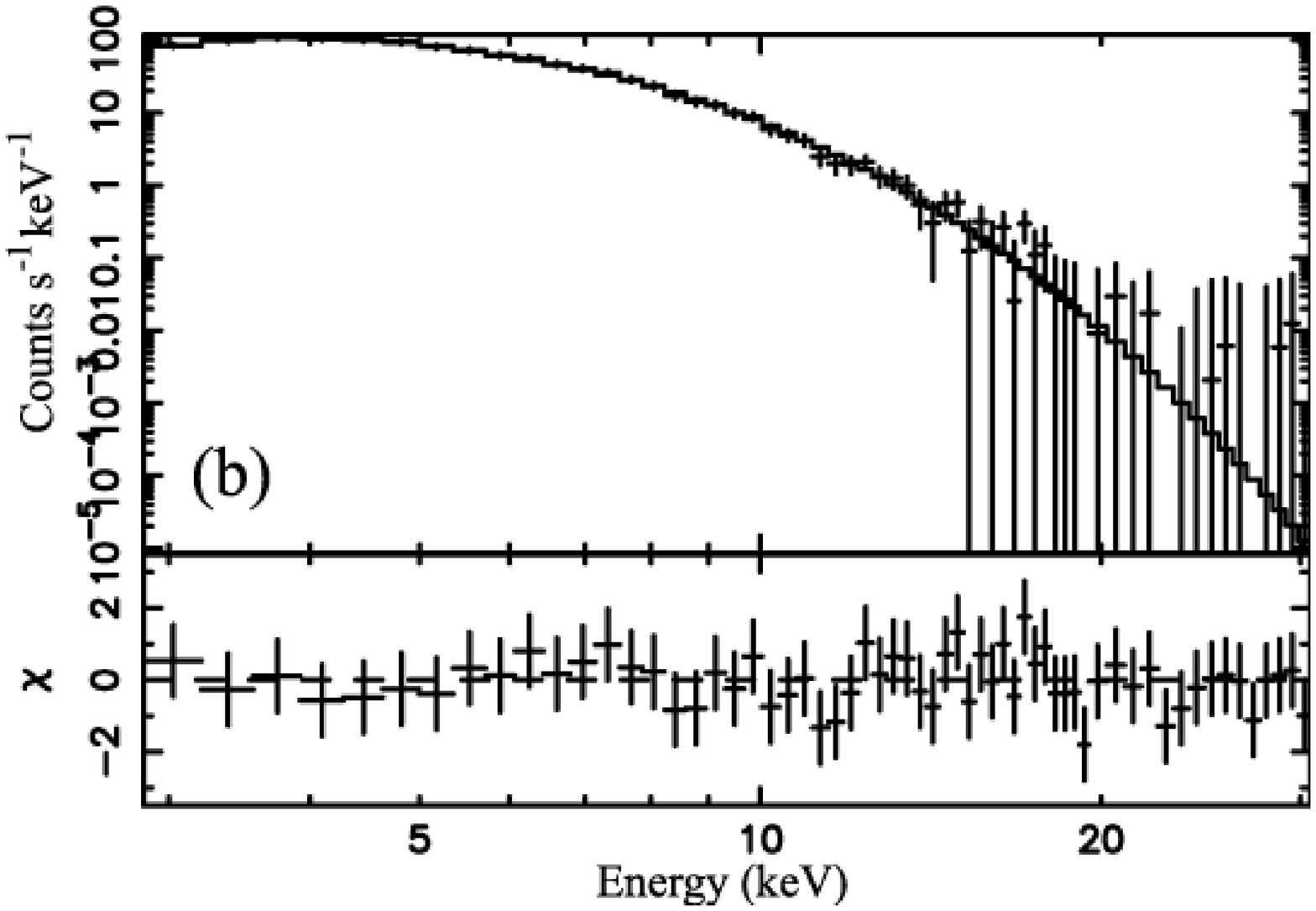}
\caption
{The same as Figure~\ref{fig:spec_1608_upper2}, 
but for the difference (Spec 2) between the faintest spectrum (Spec D) 
and the second faintest one (Spec C) in Figure~\ref{fig:spec_1608_upper1}.
It is reproduced by either a BB model (panel a, $kT_{\rm BB} \sim$ 1.3 keV) 
or an MCD model (panel b, $kT_{\rm in} \sim$ 1.7 keV), 
with the absorption column density left free.
}
\label{fig:spec_1608_upper3}
\end{center}
\end{figure}

Contrary to the above case, the other two spectra,
{the faintest spectrum (Spec D) and the second faintest one (Spec C)
which lie on a pair of  stripes adjacent to each other}, 
differ mainly in the soft energy band, 
with the hard energy end kept almost constant.
This suggests that the source variation in this case is
carried by a soft spectral component which is presumably identified
as the stable component suggested by Spec A and Spec B.
To constrain its spectral shape,
{we derived} the difference spectrum ({denoted} Spec 2) 
between Spec C and Spec D, 
and show the results in Figure~\ref{fig:spec_1608_upper3}.
{Clearly, this difference spectrum is much softer than the previous one (Spec 1), 
in agreement with the inference derived from the spectral ratios.
This Spec 2} can be reproduced by either 
an MCD model ($kT_{\rm in} \sim$ 1.7 keV) 
or a BB ($kT_{\rm BB} \sim$ 1.3 keV),
{but as shown in Table~\ref{tab:spec_1608_upper},
the former is more successful 
and gives a more reasonable absorption.}
We hence consider that the ``soft component'' is approximated 
by an MCD model with $kT_{\rm in} \sim$ 1.7 keV.

\if
The middle two spectra (Spec B and C) in Figure~\ref{fig:spec_1608_upper1} have relatively similar shapes, although their intensities are different.
As shown in Figure~\ref{fig:spec_1608_upper4}, their difference spectrum (Spec 3) is in fact softer than Spec 1, while harder than Spec 2.
If we employ the BB or MCD model to fit the spectrum, the temperatures are obtained as $kT_{\rm BB} \sim$ 2.1 keV or $kT_{\rm in} \sim$ 3.0 keV, respectively; these values are somewhat different from those derived with Spec 1 and Spec 2 (Table~\ref{tab:spec_1608_upper}), although the fits are both acceptable.
Then, we may adopt a composite model which consists of the soft MCD and the hard BB models, hereafter MCD+BB model, with $kT_{\rm in}$ fixed at 1.7 keV and $kT_{\rm BB}$ fixed at 2.5 keV, but the component normalizations are left free to vary.
We subject the two components to a common absorption.
This two component model has been found to successfully represent the difference spectrum as well.
The estimated absorption column density is roughly consistent with $0.8 \times 10^{22}$ cm$^{-2}$.
The result obtained with the BB modeling is consistent with a report by \citet{makishima_gx3+1}; namely, if employing a single BB model to represent the difference spectrum at the time when the intensity varies simultaneously all over the energy band, the BB temperature turns out to be the intermediate between those of the soft MCD model of Spec 2 and the hard BB of Spec 1.

The results obtained so far are all consistent with the Eastern model, 
where the spectra consist of the soft MCD and the hard BB components.
Since the variability of only the MCD component (Spec B) is first revealed, 
it is thought that the behavior is rarer than that of the BB.
The too high an absorption value of Spec 1 may be explained away if the soft MCD component also varies slightly between the two brightest spectra.
To confirm these inferences, we applied the Eastern model to the original four energy spectra in Figure~\ref{fig:spec_1608_upper1}, with the absorption column density fixed at $0.8 \times 10^{22}$ cm$^{-2}$.
We added a narrow Gaussian component, to represent an Fe-K emission line at $\sim$ 6.6 keV.
Because of the limited energy resolution of the PCA, the center energy and the width of the Gaussian model were constrained to fall in the range of 6.4--6.9 keV and $\le 0.2$ keV, respectively.
\fi

\begin{figure}[htbp]
\begin{center}
\includegraphics[width={0.55\textwidth}]
{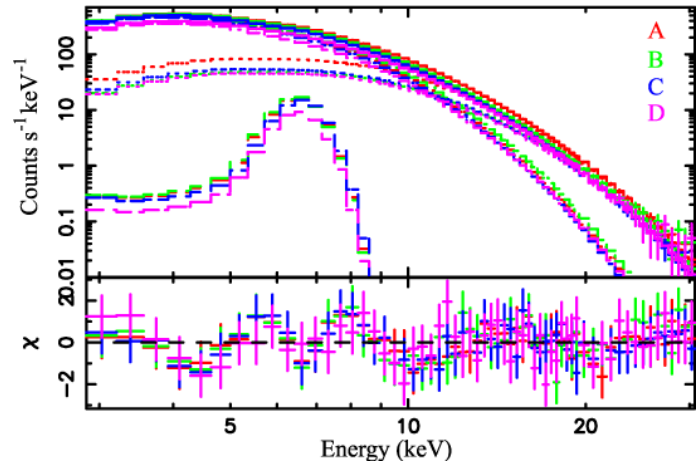}
\caption
{The same four spectra as presented in Figure~\ref{fig:spec_1608_upper1} (a), 
fitted by an MCD+BB+Gau model 
with the absorption column density fixed at $0.8 \times 10^{22}$ cm$^{-2}$.
}
\label{fig:spec_1608_upper5}
\end{center}
\end{figure}

\green{Since the above result supports the  Eastern model,
the four spectra must be reproduced} by a linear 
combination (though with different weights) of
the soft component {expressed by}  a $kT_{\rm in} \sim$ 1.7 keV MCD,
and the hard component identifiable with a $kT_{\rm BB} \sim$ 2.5 keV BB.
Then, adding a narrow Gaussian component 
to represent an Fe-K line emission line at 6--7 keV,
we fitted the four spectra with this composite model, denoted MCD+BB+Gau model.
The results are shown in Figure~\ref{fig:spec_1608_upper5} 
and Table~\ref{tab:spec_1608_upper2};
there, $r_{\rm in}$ is the innermost radius of the accretion disk 
for an assumed inclination angle $i = 0\arcdeg$,
and $r_{\rm BB}$ is that of the BB model assuming a spherical emission region.
Thus, all the four spectra have been fitted successfully by the model,
and the obtained values of $kT_{\rm in} \sim$ 1.8 keV 
and $kT_{\rm BB} \sim$ 2.7 keV indeed agree with those obtained by the two difference spectra.
These two temperatures are also consistent with those 
{found in previous works that employed}
the Eastern model \citep{mitsuda_z,makishima_gx3+1}.

Can we explain these results 
\red{
\green{with alternative  modeling?
For example,  a  model proposed by \citet{white_west},
often called  ``Western'' model,
assumed}  that Thomson opacity dominates free-free opacity in the accretion disk and the emission is significantly modified by electron scattering,
\green{while} the emission from the NS surface are hardly modified 
and observed as \green{a} simple BB emission.
Then,} an unsaturated Comptonized emission 
\green{is expected to dominate} all over the energy band, 
while a soft BB model is added in the intermediate range.
\green{This} model can actually explain the shape of Spec 2
 in terms of \green{variations} of the soft BB component.
However, Spec 1 is very difficult to explain with \green{this} model,
because its shape is different from those of either component of the model;
we would need  extreme fine tunings among the parameters of the BB 
and Comptonized components.

\blue{Returning to the Eastern model, a simple simulation was performed
to examine the increased absorption issue in Figure~\ref{fig:spec_1608_upper2}.}
\green{Employing the best fit model to Spec A, we created a fake spectrum
to be called Spec A'.
Yet  another spectrum,  named Spec B', was created,
in which the BB normalization was reduced to 67\% of that of Spec A',
and the MCD normalization was set 5\% higher than that in Spec A'.
These are meant to emulate Spec B, within the allowed fit errors 
in Table Table~\ref{tab:spec_1608_upper}.
Then, the difference between Spec A' and Spec B' was
indeed fitted successfully by a single BB model with 
nearly the same BB temperature as assumed in the input,
but with the absorption increased to $\sim 3 \times 10^{23}$ cm$^{-2}$.
}

\subsection{Absolute Values of the Physical Parameters}
\label{subsec:abs}

\green{The Eastern model allows us to assign clear physical meanings to} 
absolute values of the physical parameters obtained by the MCD and BB components.
For that purpose, however, we need to convert the measured color temperature 
$kT_{\rm X}$ (${\rm X}$ = ``in'' or ``BB'') to the effective temperature $kT^{\rm eff}_{\rm X}$, 
using so-called hardening factor $\kappa$ as
\begin{equation}
kT^{\rm eff}_{\rm X} = \kappa^{-1}~kT_{\rm X} ~.
\label{eq:T_eff}
\end{equation}
Then, the true radius $r^{\rm eff}_{\rm X}$ is estimated from the measured value $r_{\rm X}$ as
\begin{equation}
r^{\rm eff}_{\rm X} = \kappa^{2}~r_{\rm X} ~.
\label{eq:r_eff}
\end{equation}
The value of $\kappa$ is numerically estimated as 1.4--1.6 for Type I bursts \citep{ebisuzaki_hard}, 
and 1.7--2.0 for accretion disks \citep{shimura_hard}.

If we adopt $\kappa = 1.7$, the observed soft-component parameters of 
$kT_{\rm in} \sim$ 1.8 keV and $r_{\rm in} \sim 4$  km yields 
$kT^{\rm eff}_{\rm in} \sim$ 1.1 keV and $r^{\rm eff}_{\rm in} \sim$ 12 km.
Thus, the estimated true radius becomes larger than the NS radius of 10 km.
Moreover, the value of $r^{\rm eff}_{\rm in}$ is close to the radius of 
last stable orbit in terms of general relativity, $3 R_{\rm s}$ = 12.4 km,
 where $R_{\rm s} \equiv 2GM/c^{2}$ is the Schwarzschild radius, 
 $G$ is the gravitational constant, $M$ is the NS mass, and $c$ is the light speed.
As to the BB component, 
$kT_{\rm BB} \sim$ 2.7 keV and $r_{\rm BB} \sim$  1.2 km yields 
$kT^{\rm eff}_{\rm BB} \sim$ 1.8 keV and $r^{\rm eff}_{\rm BB} \sim$ 2.7 km, 
assuming $\kappa \sim$ 1.5
\green{\citep{london86, ebisu87}.}
The estimated $kT^{\rm eff}_{\rm BB}$ is close to the 
local Eddington temperature (2.0 keV) of a 1.4 $M_{\odot}$ NS, 
suggesting that the BB emission arises from a region on the NS surface.
In addition, $r^{\rm eff}_{\rm BB}$ is smaller than 10 km, 
when assuming isotropic emission from a spherical source. 
Therefore, as previously suggested by \citet{mitsuda_z},
the BB component can be regarded as being emitted from 
an equatorial zone of the NS, where the accretion disk contacts the surface.


\subsection{{Relations among the} Model Parameters} 
\label{sec:relation_mdot}

From the above spectral analysis, 
it has been confirmed that the Eastern (MCD+BB+Gau) model  successfully 
reproduces the {selected four} spectra in the upper-banana state,
and yields physically reasonable interpretations.
{Given these,} we applied the MCD+BB+Gau model to the 95 spectra in the upper-banana state, 
and obtained $\chi^2$/d.o.f. of $< 1.4$ for all of them.
{Therefore, we regard the Eastern model
as applicable to all the present data sets.}

\begin{figure}[htbp]
\begin{center}
\includegraphics[width={0.55\textwidth}]
{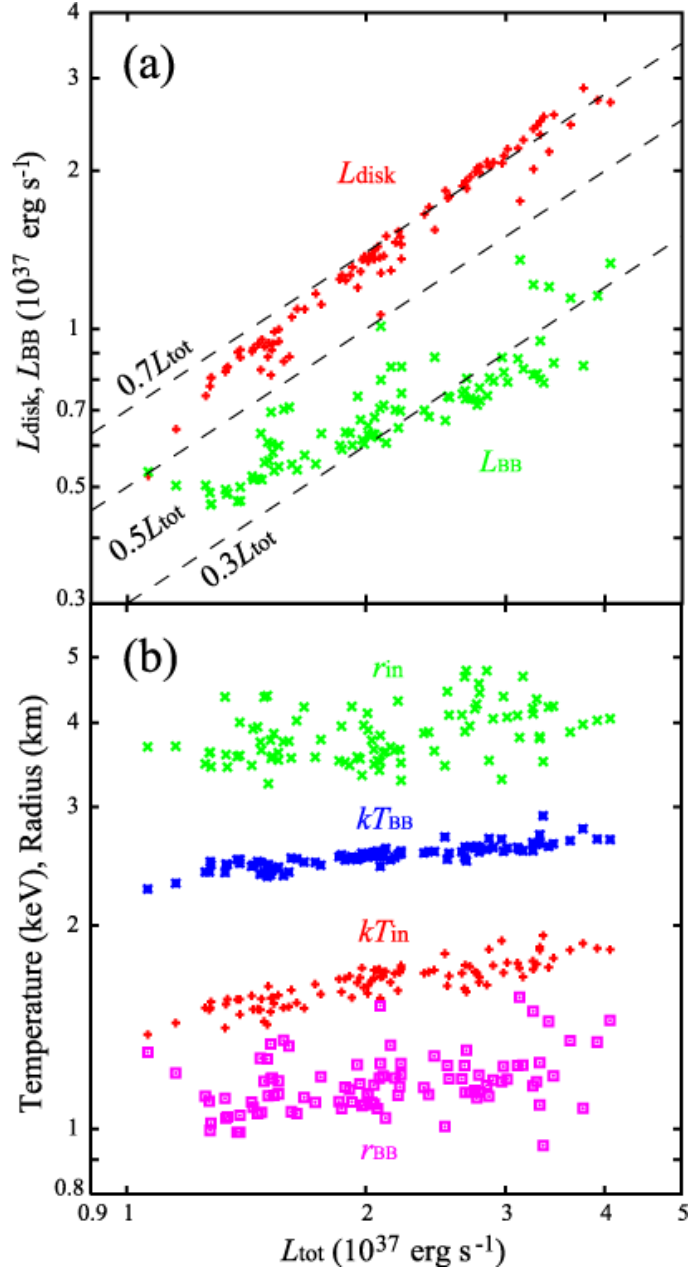}
\caption
{
(a) The relations of the MCD luminosity ($L_{\rm disk}$) and the BB luminosity ($L_{\rm BB}$)
as a function of $L_{\rm tot} \equiv L_{\rm disk} + L_{\rm BB}$ in the upper-banana state.
Each data point represents one of the 95 observations.
The three dashed lines show
$0.7 L_{\rm tot}$, $0.5 L_{\rm tot}$, and $0.3 L_{\rm tot}$, from top to bottom.
(b) The same as (a) but for
the temperatures and radii of the MCD and BB models.
}
\label{fig:spec_1608_all1}
\end{center}
\end{figure}

Figure~\ref{fig:spec_1608_all1} summarizes the relation 
among the \green{unabsorbed} disk bolometric luminosity $L_{\rm disk}$, 
the  \green{unabsorbed} BB bolometric luminosity for isotropic emission $L_{\rm BB}$, 
and their sum $L_{\rm tot} \equiv L_{\rm disk} + L_{\rm BB}$,
as well as the temperatures and radii.
Here, $L_{\rm disk}$ is calculated assuming a face-on disk 
(i.e., the inclination angle $i = 0\arcdeg$) as
\begin{equation}
L_{\rm disk} = 2 \times \int^{\infty}_{r_{\rm in}} 2\pi{r}\sigma T(r)^{4} dr = 4\pi r_{\rm in}^{2} \sigma T_{\rm in}^{4} ~, \label{eq:lbol_mcd}
\end{equation}
where the first factor of 2 means emission from the two sides of the disk, 
$\sigma$ is the Stefan-Boltzmann constant, 
and $T(r)$ is the disk temperature at the radius $r$.

In Figures~\ref{fig:spec_1608_all1} (a),
both $L_{\rm disk}$ and $L_{\rm BB}$ are seen to increase 
as $L_{\rm tot}$ increases from $1 \times 10^{37}$ to $4 \times 10^{37}$ erg s$^{-1}$.
However, a closer inspection reveals that $L_{\rm disk}$ varies more steeply than $L_{\rm BB}$.
Actually, the data behavior in Figure~\ref{fig:spec_1608_all1} (a) can be approximated as
\begin{equation}
 L_{\rm disk} \propto L_{\rm tot}^{1.1},~~~~~L_{\rm BB} \propto L_{\rm tot}^{0.7},~~~~~L_{\rm BB} \propto L_{\rm disk}^{0.6} ~. \label{eq:1608_Ltot_Ldisk_Lbb}
\end{equation}
As a result, the ratio between $L_{\rm BB}$ and $L_{\rm disk}$ decreases from 0.6 to 0.4 as $L_{\rm tot}$ increases by a factor of 4,
 in agreement with previous reports \citep{gilfanov_vari,revnivtsev_vari}.
Similarly, Figure~\ref{fig:spec_1608_all1} (b) shows the luminosity dependences of the temperature and radius parameters,
derived in the same way as in \S~3.1,
and presented without applying any of the correction factors mentioned in \S~3.2.
The relations are approximated as
\begin{equation}
kT_{\rm in} \propto L_{\rm tot}^{0.19},~~~~~
r_{\rm in} \propto L_{\rm tot}^{0.12},~~~~~
kT_{\rm BB} \propto L_{\rm tot}^{0.10},~~~~~
r_{\rm BB} \propto L_{\rm tot}^{0.11} ~. \label{eq:1608_Ltot_t_r}
\end{equation}

\begin{figure}[htbp]
\begin{center}
\includegraphics[width={0.55\textwidth}]
{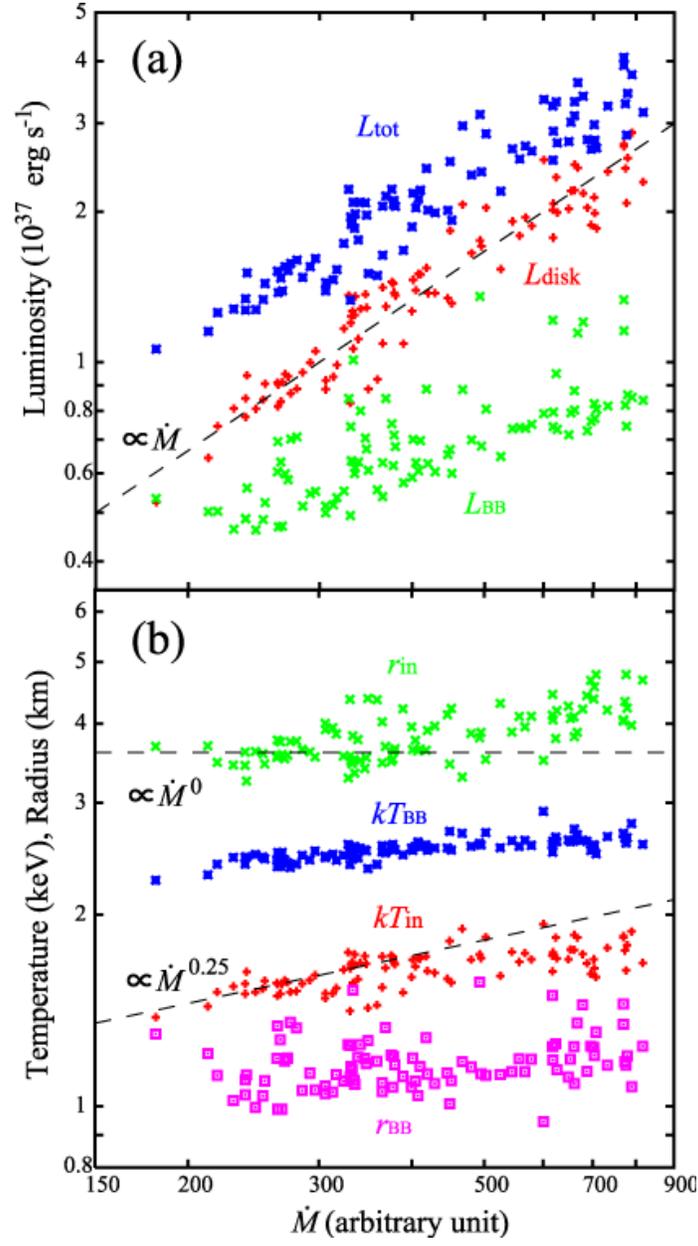}
\caption
{
The same as Figure~\ref{fig:spec_1608_all1},
but shown as a function of the estimated mass accretion rate $\dot{M}$.
The three dashed lines illustrate dependences as 
$L_{\rm disk} \propto \dot{M}$, $r_{\rm in} \propto \dot{M}^{0}$, 
and $kT_{\rm in} \propto \dot{M}^{0.25}$.
}
\label{fig:spec_1608_all3}
\end{center}
\end{figure}

Instead of $L_{\rm tot}$, we may utilize the mass accretion rate $\dot{M}$,
which can be estimated in the following way.
Generally, the accreting matter releases energy at a rate of $GM\dot{M}/r_{\rm in}$, as it flows through the disk down to the innermost radius $r_{\rm in}$.
According to the picture of the standard accretion disk,
a half of this luminosity is radiated away in the accretion disk 
as $L_{\rm disk} = \frac{1}{2} \frac{GM\dot{M}}{r_{\rm in}}$,
and the other half is stored in the Keplerian kinetic energy.
Therefore, the mass accretion rate is estimated as
\begin{equation}
 \dot{M} = \frac{2 L_{\rm disk} r_{\rm in}}{GM} \propto L_{\rm disk} r_{\rm in} \propto r_{\rm in}^{3} T_{\rm in}^{4}.
\label{eq:mdot}
\end{equation}
This relation remains valid even if $r_{\rm in}$ varies, 
as long as the disk stays in the standard state.
Figure~\ref{fig:spec_1608_all3} shows the same physical parameters
as presented in Figure~\ref{fig:spec_1608_all1},
but this time, as a function of $\dot{M}$ estimated via equation~(\ref{eq:mdot}),
where the unit of $\dot{M}$ is arbitrary.

In the case of a standard accretion disk, 
$L_{\rm disk}$ and $kT_{\rm in}$ are expected to be proportion to $\dot{M}$ and $\dot{M}^{0.25}$, 
respectively \citep[e.g., ][]{makishima_bhb,ebisawa_lmcx-3,tanaka_novae},
as long as $r_{\rm in}$ kept constant.
Indeed, in the range of $\dot{M} \lesssim 300$
\red{(\green{corresponding to} $L_{\rm tot} \lesssim 1.5 \times 10^{37}$ erg s$^{-1}$)},
Figure~\ref{fig:spec_1608_all3} reveals tight scalings as
\begin{equation}
L_{\rm disk} \propto \dot{M}^{0.92 \pm 0.07},
~~~~~kT_{\rm in} \propto \dot{M}^{0.21 \pm 0.05},
~~~~~r_{\rm in} \propto \dot{M}^{0.05 \pm 0.07}~, \label{eq:1608_Mdot_disk1}
\end{equation}
which agree, within errors, with the predictions for standard disks.
This result is consistent with that reported by \citet{lin_aql1608}.

As the accretion rate $\dot{M}$ increases to $\gtrsim 300$ in Figure~\ref{fig:spec_1608_all3},
the disk parameters start deviating from the scalings of equation~(\ref{eq:1608_Mdot_disk1}), 
and follow new relations approximated as
\begin{equation}
L_{\rm disk} \propto \dot{M}^{0.80 \pm 0.03},
~~~~~kT_{\rm in} \propto \dot{M}^{0.11 \pm 0.02},
~~~~~r_{\rm in} \propto \dot{M}^{0.18 \pm 0.03}~. \label{eq:1608_Mdot_disk2}
\end{equation}
Thus, the inner disk radius $r_{\rm in}$ apparently starts ``retreating'',
with increasing fluctuations,
but the increase in $r_{\rm in}$ is compensated by the flattering in $kT_{\rm in}$.
{As a result,  $L_{\rm disk}$ starts to weakly saturate,
instead of increasing in proportion to $\dot{M}$.}

\if
This behavior is thought to be consistent with results by \citet{popham_boundary}, who numerically calculated the physical condition of the boundary layer and estimated that the innermost radius of the accretion disk retreats back by the increased inner-radiation pressure when the mass accretion rate approaches the Eddington limit.
Indeed, the number of the $kT_{\rm in}$ index $\sim 0.11$ is naturally understood based on the scheme of the standard accretion disk model with the increasing innermost radius.
\fi
\if
The standard disk has a radial relation of $T(r) \propto r^{-0.75}$, where $T(r)$ is the temperature at the radius $r$, and a mass-accretion-rate relation of $T(r) \propto \dot{M}^{0.25}$ at the same radius.
Then, the increase of the innermost temperature is modified as,
\begin{equation}
kT_{\rm in} \propto \dot{M}^{0.25}~r_{\rm in}^{-0.75} \propto \dot{M}^{0.25}~(\dot{M}^{0.18})^{-0.75} \propto \dot{M}^{0.11}~. \label{eq:1608_Mdot_disk3}
\end{equation}
Therefore, $kT_{\rm in}$ is thought to represent the temperature of the standard disk at the retreated radius $r_{\rm in}$.
\fi

Turning to the BB parameters, we find $kT_{\rm BB}$ rather stable, 
with a weak positive dependence on $\dot{M}$.
On the contrary, $r_{\rm BB}$ exhibits a large scatter, 
which causes similar scatter in $L_{\rm BB}$.
The $\dot{M}$-dependence of these BB parameters can be approximated as
\begin{equation}
L_{\rm BB} \propto \dot{M}^{0.5},
~~~~~kT_{\rm BB} \propto \dot{M}^{0.07 \pm 0.01},
~~~~~r_{\rm BB} \propto \dot{M}^{0.13}~. \label{eq:1608_Mdot_bb}
\end{equation}
This BB behavior is the same {as previously reported
\citep{mitsuda_z,gilfanov_vari,revnivtsev_vari}.}
We also find that $kT_{\rm BB}$ is always higher than $kT_{\rm in}$, and $r_{\rm BB}$ is smaller than $r_{\rm in}$.
These relations are consistent with the {basic assumptions} of the Eastern model;
the BB emission comes from the NS surface, 
while the MCD component from the surrounding accretion disk.

\subsection{Effective Degrees of Freedom Causing Spectral Variability}
\label{subsec:dof}

We have so far described the energy spectra and their variations
in terms of the four quantities,
$kT_{\rm in}$, $r_{\rm in}$, $kT_{\rm BB}$, and $r_{\rm BB}$.
However, it is not yet clear how many of them can be regarded
as independent variables (with the rest depending on them).
In other words, we need to know
how many degrees of freedom are involved
in the observed spectral variations in the upper-banana state.
This can be done by applying  ``fractal dimension analysis''
\citep[e.g., ][]{yukari_dron}
to the four model parameters.

For this purpose, let us define a 4-dimensional vector space
spanned by the four variables,
$Y_1=kT_{\rm in}$, $Y_2=r_{\rm in}$,
$Y_3=kT_{\rm BB}$, and $Y_4=r_{\rm BB}$,
wherein each data set is represented by a vector
\[
\vec{V(i)}=\left\{Y_1(i), ~Y_2(i), ~Y_3(i),~ Y_4(i) \right\}
\]
with  $i=1, 2, .., 95$ denoting the data number.
Let us also define the average vector
$\langle \vec{V} \rangle = \Sigma_{i=1}^{95} \vec{V(i)}/95
 \equiv \left\{\langle Y_1 \rangle, \langle Y_2 \rangle,
 \langle Y_3 \rangle, \langle Y_4 \rangle \right\}$,
and the zero-mean vectors,
 $\vec{v(i)} = \vec{V(i)} - \langle \vec{V} \rangle
 \equiv \left\{y_1(i), y_2(i), y_3(i), y_4(i) \right\}$
 {with $y_k(i) \equiv Y_k(i) -\langle Y_k \rangle$.}
Then, a ``distance" $D(i)$ of the  vector $\vec{v(i)}$,
measured from the origin, is calculated as
 \begin{equation}
D(i) = \left[ \sum_{k=1}^{4}\left\{y_k(i)/\sigma_k \right\}^2 \right]^{1/2}~ (i = 1, 2, \cdots, 95),
\label{eq:distance}
\end{equation}
{
where
%
 $\sigma_k\ = \left[ \sum_{i=1}^{95} y_k(i)^2 /94 \right]^{1/2}$
is the standard deviation in $y_k(i)$ around the origin
[or in $Y_k(i)$ around its mean $\langle Y_k \rangle$].
}
Finally, we calculate the number $N(<D)$ of those data points
of which the distance $D(i)$
is less than a given value $D$.

\begin{figure}[htbp]
\begin{center}
\includegraphics[width={0.55\textwidth}]
{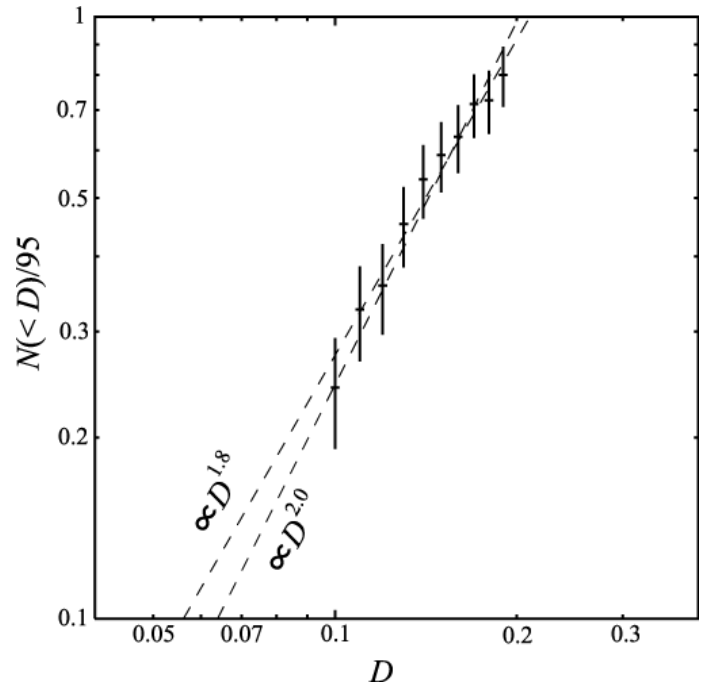}
\caption
{
Results of the fractal dimension analysis of the four model parameters
($kT_{\rm in}$, $r_{\rm in}$, $kT_{\rm BB}$, and $r_{\rm BB}$)
over the 95 data sets.
See text for details.
}
\label{fig:fractal}
\end{center}
\end{figure}

Figure~\ref{fig:fractal} shows the normalized data point number
$N(<D)/95$ from equation~(\ref{eq:distance}) as a function of $D$,
over the range of $N/95 = 0.2-0.8$ (or $N =23-76$).
If the variations are controlled by $n~(1\le n \le4) $ independent parameters,
we expect the vectors $\vec{v(1)}, \vec{v(1)}, ... \vec{v(95)}$ to form
a $n$-dimensional subspace in the vector space,
so that $N(<D)$ should increase as $\propto D^n$.
Indeed, Figure~\ref{fig:fractal} reveals a tight power-law relation  as
\begin{equation}
N \propto D^{1.8 \pm 0.1}~.
\label{eq:fractal}
\end{equation}
Since this {result is  consistent} with  $n=2$,
we infer that the spectral behavior in the upper-banana state of 4U 1608--522
has effectively two degrees of freedom.

\subsection{Fluctuations Independent of the Total Luminosity}
\label{subsec:vari}

Of the two independent variables
describing the spectral variability  (\S~\ref{subsec:dof}),
one is obviously the total  luminosity $L_{\rm tot}$,
or nearly equivalently, the mass accretion rate $\dot{M}$.
Actually in Figure 6 (or Figure 7),
the four quantities are all observed to
depend primarily on $L_{\rm tot}$ (or $\dot{M}$).
However, they  show significant scatters
around the $L_{\rm tot}$-dependent {correlation},
so that none of them can be regarded as a single-valued function
of $L_{\rm tot}$ (or $\dot{M}$).
This is consistent with the presence of
the second degree of freedom {revealed in \S~\ref{subsec:dof}}.
In order to identify what is causing this extra freedom,
we remove  the $L_{\rm tot}$-dependence
from the behavior of the four model parameters,
and study the residual variations.

\begin{figure}[htbp]
\begin{center}
\includegraphics[width={0.55\textwidth}]
{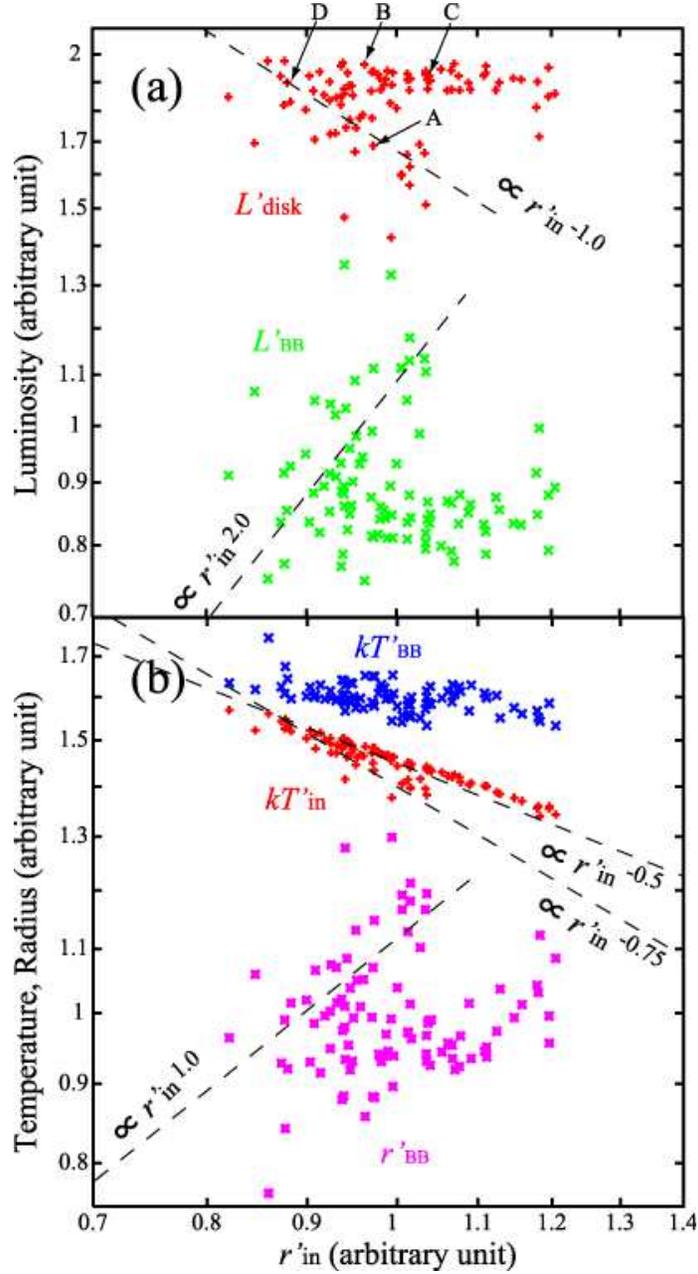}
\caption
{
The same as Figure~\ref{fig:spec_1608_all1},
but for the physical parameters
from which the $L_{\rm tot}$ dependence was removed via equations~(12)--(\ref{eq:detrend}).
They are plotted as a function of the de-trended innermost radius
of the accretion disk $r'_{\rm in}$.
Dashed lines indicate the relations of $r'_{\rm in}$$^{p}$,
with $p = -1.0, -0.75, -0.5$, 1.0 and 2.0.
The $L'_{\rm disk}$ parameters of Spec A to D in Table~\ref{tab:detrend} are also indicated.
}
\label{fig:detrend}
\end{center}
\end{figure}

To eliminate the {major $L_{\rm tot}$-dependence} from the physical parameters,
we calculated their de-trended counterparts {$Z'(i)$
($Z = L_{\rm disk}, L_{\rm BB}, kT_{\rm in}, r_{\rm in}, kT_{\rm BB}$ and $r_{\rm BB}$), }
employing equations~(\ref{eq:1608_Ltot_Ldisk_Lbb}) and (\ref{eq:1608_Ltot_t_r}), as
\begin{eqnarray}
 L'_{\rm disk}(i) \propto L_{\rm disk}(i)/L_{\rm tot}(i)^{1.1},~~~~~
 L'_{\rm BB}(i)  \propto L_{\rm BB}(i)/L_{\rm tot}(i)^{0.7}, \\
 kT'_{\rm in}(i) \propto kT_{\rm in}(i)/L_{\rm tot}(i)^{0.19},~~~~~
 r'_{\rm in}(i)  \propto r_{\rm in}(i)/L_{\rm tot}(i)^{0.12}, \\
 kT'_{\rm BB}(i) \propto kT_{\rm BB}(i)/L_{\rm tot}(i)^{0.10},~~{\rm{and}}~~~
 r'_{\rm BB}(i)  \propto r_{\rm BB}(i)/L_{\rm tot}(i)^{0.11} ~. \label{eq:detrend}
\end{eqnarray}
\green{(As the de-trending parameter, we chose $L_{\rm tot}$ rather than $\dot{M}$,
because the latter is less directly estimated from the data.)}
Absolute values of the de-trended parameters are not meaningful,
and we hereafter consider their relative values only.
Since $r'_{\rm in}$ shows the largest variation among them,
we plot {in Figure~\ref{fig:detrend}} the de-trended parameters 
as a function of $r'_{\rm in}$.
Table~\ref{tab:detrend} also summarizes those of Spec A to D in $\S$~\ref{subsec:spec}.
From the behavior of the de-trended luminosities,
the data points can be divided into two branches;
one has almost constant luminosities as $r'_{\rm in}$ varies,
while the other is characterized by significant $r'_{\rm in}$-dependent variations 
both in $L'_{\rm disk}$ and $L'_{\rm BB}$.
Hereafter, the two branches are denoted as
``constant-luminosity branch (CLB)'' and ``variable-luminosity branch (VLB)'', respectively.
The two branches may connect at $r'_{\rm in} \sim 0.85$, 
rather than behaving independently from each other.

In the CLB, the  de-trended BB luminosity  stays nearly constant at  $L'_{\rm BB} \sim 0.8$;
so are  the two BB parameters, $kT'_{\rm BB} \sim 1.6$ and $r'_{\rm BB} \sim 0.9$.
Similarly, the de-trended disk luminosity  remains at $L'_{\rm disk} \sim 0.9$.
As a result, the CLB data points are distributed in Figure~\ref{fig:spec_1608_all1} (a)
along the major correlation trends.
Nevertheless, $r'_{\rm in}$ varies by $\sim \pm 20$\%,
accompanied by a clear decrease in the de-trended disk temperature
as  $kT'_{\rm in} \propto r_{\rm in} $$^{-0.5}$.
(This scaling is a natural consequence of the constant $L'_{\rm disk}$
and the relation of  $L'_{\rm disk} \propto r'_{\rm in}$$^{2} T'_{\rm in}$$^{4}$.)
There are two possibilities to explain this $r'_{\rm in}$ behavior.
One is real changes in $r_{\rm in}$,
and the other is those in the color hardening factor $\kappa$ of the disk.
Since $L'_{\rm disk}$ is kept constant in the CLB,
the latter case may be more likely.
\green{
Specifically, the observed behavior will be explained
if some unspecified mechanisms 
(possibly related to vertical structure changes in the disk) 
caused $\kappa$ to increased by $\sim 10\%$
while somehow keeping the effective temperature of the disk and its true radius constant;
\blue{the} observed color temperature $kT'_{\rm in}$ would then 
increase by $\sim 10\%$ according to equation~(\ref{eq:T_eff}),
and the apparent disk radius $r'_{\rm in}$ would decrease 
by $\sim 20\%$ after eqation~(\ref{eq:r_eff}),
\blue{thus reproducing the observaton.}
}

In the VLB,  $r'_{\rm in}$ varies over a relatively small  range ($\sim \pm 10$\%),
whereas the two luminosities both vary significantly in an anti-correlated way;
$L'_{\rm disk}$ decreases from 1.9 to 1.5 as $\propto r'_{\rm in}$$^{-1}$,
while $L'_{\rm BB}$ increases from 0.8 to 1.2 as $\propto r'_{\rm in}$$^{2}$.
This complementary behavior between the MCD and BB components
appears in Figure~\ref{fig:spec_1608_all1} 
as the large fluctuations away from  the major trend,
where the data points  reach a level of $L_{\rm disk} \sim L_{\rm BB} \sim 0.5 L_{\rm tot}$.
\green{(This behavior cannot be a fitting artifact
caused by strong couplings between the two spectral component,
since  the nominal fitting errors are not particularly larger in the VLB,
and are generally smaller than/comparable to the size of the plotting symbols
in Figure~\ref{fig:spec_1608_all1} and Figure~\ref{fig:spec_1608_all3}.)
}
Since $kT'_{\rm BB}$ is almost constant (or only slightly decreasing) like in the CLB,
the increase of $L'_{\rm BB}$ is mainly caused by that of $r'_{\rm BB}$ 
from 0.9 to 1.2  as  $\propto r'_{\rm in}$$^{1}$.
We again observe $kT'_{\rm in}$  to decrease like in the CLB,
but  with a different  $r'_{\rm in}$-dependence
which is approximated as  $\propto r'_{\rm in}$$^{-0.75}$.
This is exactly the behavior of a standard accretion disk \citep{shakura_disk}
when its radius varies under a constant $\dot{M}$,
because we then expect $\dot{M} \propto r_{\rm in}^3 T_{\rm in}^4$ (equation~(\ref{eq:mdot})).
This also explains the observed disk luminosity behavior, 
as  $L'_{\rm disk} \propto r_{\rm in}^{2}T_{\rm in}$$^{4}  \propto r'_{\rm in}$$^{-1}$.
\green{Obviously, the increase in the disk radius under a constant $\dot{M}$
predicts that a larger fraction of the overall gravitational energy release
should be emitted by the BB component;
in fact, along the full VLB in Figure~\ref{fig:spec_1608_all3}(b),
$L'_{\rm disk}$ is seen to decrease 
\blue{by $\sim 0.3$ in the employed arbitrary unit,
while $L'_{\rm BB}$ to increase by $\sim 0.3$, thus conserving the total luminosity.}}
In other words, the VLB is characterized by actual changes in the innermost disk radius.

\if
On the other hand, when $r'_{\rm in}$ varies in relatively small range ($\sim \pm 10$\%),
the VLB changes the amplitudes of both luminosities significantly
in the anti-correlated way;
$L'_{\rm disk}$ decreases from 1.9 to 1.5 with $\propto r'_{\rm in}$$^{-1}$
while $L'_{\rm BB}$ increases from 0.8 to 1.2 with $\propto r'_{\rm in}$$^{2}$.
This complementary behavior between the MCD and BB components
corresponds to the large fluctuation from the major trend in Figure~\ref{fig:spec_1608_all1}
(e.g., the datasets with $L_{\rm disk} \sim L_{\rm BB} \sim 0.5 L_{\rm tot}$).
Since $kT'_{\rm BB}$ is almost constant (or slightly decreasing) similarly to the CLB,
the increase of $L'_{\rm BB}$ is mainly caused by that of $r'_{\rm BB}$ 
from 0.9 to 1.2 close to $\propto r'_{\rm in}$$^{1}$.
The value of $kT'_{\rm in}$ again decreases as well as in the CLB,
but the $r'_{\rm in}$ dependence is relatively strong
and nearly $\propto r'_{\rm in}$$^{-0.75}$,
which is the relation of the standard accretion disk \citep{shakura_disk}
and results in $L'_{\rm disk} \propto r'_{\rm in}$$^{-1}$.
\fi

Based on these correlation analyses among the de-trended spectral parameters,
we conclude that the second independent variable (besides $\dot{M}$; \S~\ref{subsec:dof})
causing the variability in the upper-banana state can be
ascribed to sporadic changes in $r'_{\rm in}$ (or in $r_{\rm in}$).
However, this $r'_{\rm in}$ variability is likely to be 
further subdivided into
two different mechanisms under a constant $\dot{M}$;
one is an apparent effect due to variations in $\kappa$ of the disk,
and the other is real changes.

As shown in Figure~\ref{fig:detrend} and Table~\ref{tab:detrend},
the $L'_{\rm disk}$ parameters of Spec B, C and D locate on the CLB line,
while that of Spec A does on the VLB one.
Therefore, we consider that the hard/soft-band differences of Spec A/B and C/D observed in Figure~\ref{fig:spec_1608_upper1} are mainly attributed to the VLB and CLB variation, respectively.
On the other hand, the de-trended parameters of Spec B and C are similar,
and the variability between them observed over the entire 3--30 keV band
is thought to be dominated by the $\dot{M}$ change.


\section{DISCUSSION AND CONCLUSION}

\subsection{{The Eastern Model}}

Through the analysis of the {\it RXTE} spectra of the atoll source 
4U 1608--522 in its upper-banana state, we have confirmed
that both time-averaged and difference energy spectra
can be reproduced successfully by the Eastern (MCD+BB+Gau) model.
Considering the hardening factor,
the effective temperature and radius of the MCD component were obtained 
as $kT^{\rm eff}_{\rm in} \sim$ 1.1 keV and $r^{\rm eff}_{\rm in} \sim$ 12 km,
respectively.
The radius is larger than the representative NS radius, 10 km,
and is consistent with the last stable orbit, $3 R_{\rm s}$ = 12.4 km,
allowed by general relativity.
This result agrees with the picture of 
a standard accretion disk which is formed around a NS.
The BB parameters were found as
$kT^{\rm eff}_{\rm BB} \sim$ 1.8 keV and $r^{\rm eff}_{\rm BB} \sim$ 2.7 km,
assuming isotropic emission from a spherical source.
The temperature is close to the local Eddington temperature (2.0 keV) at the NS surface,
and the radius is smaller than 10 km.
Then, the BB component can be regarded as being
emitted from an equatorial zone of the NS,
as previously suggested by \citet{mitsuda_z}.

\subsection{Saturation of the Blackbody Luminosity}

According to the picture of standard accretion disks,
a half of the released gravitational energy is 
radiated from the accretion disk ($L_{\rm disk}$),
and the other half is stored in the Keplerian kinetic energy 
and then emitted ($L_{\rm BB}$)
when the matter settles onto the NS surface.
Then, we expect $L_{\rm BB}$ to be proportion to $L_{\rm disk}$,
and hence the $L_{\rm BB}/L_{\rm disk}$ ratio to be constant.
Indeed, $L_{\rm disk}$ was found to increase almost linearly
with the total luminosity $L_{\rm tot}$ (figure 6a).
However, the same figure reveals that $L_{\rm BB}$ increases less steeply,
making the $L_{\rm BB}/L_{\rm disk}$ ratio decrease from 0.6 to 0.4
as $L_{\rm tot}$ increases from $1 \times 10^{37}$ to $4 \times 10^{37}$ erg s$^{-1}$.

We may think of two possibilities to explain 
the observed relative decrease of $L_{\rm BB}$.
One is that the accreting matter reaches the NS surface
and emits $L_{\rm BB}$ which is equivalent to the Keplerian energy
 (i.e. $\propto L_{\rm disk}$),
but we cannot observe all the emission
because its wavelength shifts outside the PCA energy band (3--30 keV),
or the geometrical angle of the emission moves from our line of sight.
However, we have not observed any hint of extra emission
in the softest or hardest energy ends of the PCA spectra.
In addition, other sources,
which are thought to have different inclination angles,
are also reported to show the same behavior of 
the decreasing $L_{\rm BB}/L_{\rm disk}$ ratio \citep{revnivtsev_vari}.
Therefore, this possibility is unlikely.

The other possibility is 
that the accreting matter flows all the way through the disk 
down to its inner radius that is close to the NS surface,
but its progressively larger fraction fails to accrete onto the NS surface.
If the accretion-failed matter stayed around the NS away from the surface
(e.g., an expansion of the boundary layer),
it would accumulate to become optically thick,
eventually producing detectable emission.
This would lead to a decrease in $kT_{\rm BB}$,
because the emission radius should increase.
As shown in Figure~\ref{fig:spec_1608_all1} and equation~(\ref{eq:1608_Ltot_t_r}),
however, this is not the case;
the $kT_{\rm BB}$ value stays almost constant 
(or rather increases) as the total luminosity increases.
Then, a fraction of the matter must be outflowing and escaping from the system
without releasing its kinetic energy as emission.
\blue{Although the observed $L_{\rm tot} \sim 4 \times 10^{37}$ erg s$^{-1}$
is only $\sim 20\%$ of the  Eddington luminosity for a 1.4 $M_{\odot}$ NS,
the observed value of $kT^{\rm eff}_{\rm BB} \sim 1.8$ keV is
already close to the spherical Eddington temperature at 10 km (2.0 keV).
Considering further the non-spherical geometry of the disk,
and the disk radius which is possibly larger than 10 km,
the putative outflow is likely to be driven by
increased radiation pressure in the innermost disk region.
}
The presence of such outflows is indeed suggested
by detections of broad absorption features from LMXBs
(Schulz \& Brandt 2002; Ueda et al.\ 2004),
black hole binaries (Kotani et al.\ 2000; Yamaoka et al.\ 2001; Kubota et al.\ 2007),
and  active galactic nuclei (Pounds et al.\ 2003ab).

Assuming that the decrease in  $L_{\rm BB}/L_{\rm disk}$
from 0.6 to 0.4 is due to outflows,
about $(0.6-0.4)/0.6 \sim$ 30\% of the accreting matter
is estimated to escape from the system
at a typical luminosity of $\sim 4 \times 10^{37}$ erg s$^{-1}$.
\green{Toward lower luminosities,
the ratio appears to converge to $\sim 0.6$, 
rather than 1.0 which would be theoretically expected.
This may be attributed to effects due, e.g., 
to the system inclination, general relativity,
and NS rotation (e.g., \citet{sunyaev_ratio}).
Detailed  theoretical discussion on this point
is  beyond the scope of the present paper.}

As indicated by equation~(\ref{eq:1608_Mdot_disk2}),
the innermost disk radius $r_{\rm in}$ is observed to increase gradually
\green{when the mass accretion rate increases 
to$\gtrsim 300$ in Figure~\ref{fig:spec_1608_all3},
or equivalently, when the total luminosity becomes $ \gtrsim 1.5 \times 10^{37}$ erg s$^{-1}$
(corresponding to \blue{$\gtrsim 7\%$} of the Eddington luminosity).}
This is unlikely to be an apparent effect caused, e.g.,
by changes in the hardening factor,
since this quantity would increase
as the luminosity increases \citep{gierlinski_hard,davis_hard},
and would hence make $r_{\rm in}$ smaller.
Therefore, the increase in $r_{\rm in}$ toward higher accretion rates
is considered to represent a real retreat of the innermost disk radius.
This interpretation \red{may} agree with the results by \citet{popham_boundary},
who numerically calculated the physical condition of the boundary layer
and showed that the disk inner edge retreats back
by the increased radiation pressure
when the luminosity approaches the Eddington limit.

\if
In Figure~\ref{fig:spec_1608_all1} and \ref{fig:spec_1608_all3}, 
the increase of $L_{\rm BB}$ is mainly caused by that of $r_{\rm BB}$, 
while $kT_{\rm BB}$ {remains almost constant.}
This may be also because of the high-radiation pressure 
of $kT_{\rm BB}$ close to the Eddington temperature,
since the matter tends to escape such high-pressure regions and accrete onto other places,
expanding the emitting equatorial region on the NS surface rather than increasing the temperature.
\fi

\if
Along the increase of $\dot{M}$,
we consider that the physical state of the accretion flow will evolve in the following manner.
At the beginning of the upper-banana state ($\dot{M} \lesssim 300$),
there should be little outflows,
and all the matter accretes from the standard accretion disk
to a relatively narrow equatorial region on the NS surface.
In this stage, the BB emission comes from the narrow belt region of the NS surface.
When $\dot{M}$ increases,
the accretion disk behaves in the ``standard'' manner;
the innermost temperature $kT_{\rm in}$ is proportion to $\sim \dot{M}^{0.25}$,
and the innermost radius $r_{\rm in}$ stays almost constant.
Since $kT_{\rm BB}$ stays uniform and almost consistent with the Eddington temperature,
we then expect the belt region emitting the BB component becomes vertically wider,
and hence the observed $r_{\rm BB}$ increases.
\fi

\if
As $\dot{M}$ increases $\gtrsim 300$,
the disk may stay in the standard state,
but the transition region will become vertically thicker
due to the increased radiation pressure inside the disk.
Due to the temperature uniformity, the outward radiation pressure is expected to be latitudinally rather constant, while the inward gravitational pressure must decrease toward higher (positive and negative) latitudes due to the reduced matter density.
As a result, we expect the outflow to take place first at the ``northern'' and ``southern'' edge of the belt, and proceed toward the equatorial region.
This scenario presented above explains, at least qualitatively, the relative decrease in the $L_{\rm BB}/L_{\rm disk}$ (and hence $L_{\rm BB}/\dot{M}$) ratio.
\fi

\subsection{{Secondary Spectral Variations}}

The result of the fractal dimension analysis shows 
that the source variations are controlled by two independent variables.
One of them is obviously the mass accretion rate
(or nearly equivalently the total luminosity),
which produces the major trend in Figure~\ref{fig:spec_1608_all3}
as discussed so far.
The other variable is thought to be random changes in the
inner disk radius, as represented by the behavior of
$r'_{\rm in}$  in the de-trended analysis performed in \S~\ref{subsec:vari}.
As discussed therein, 
the variations of $r'_{\rm in}$
may be
caused by 
two different mechanisms,
corresponding to the CLB and VLB.
These two branches are considered to degenerate
in the fractal dimension analysis,
and difficult to identify as two independent degrees of freedom.

In the CLB, the change in $r'_{\rm in}$ 
seems to be
apparent rather than real,
and is considered to reflect fluctuations in the hardening factor of the disk
As a result, the BB parameters remain unchanged,
while the MCD parameters vary as  $kT'_{\rm in} \propto r'_{\rm in}$$^{-0.5}$, 
keeping the disk luminosity constant ($L'_{\rm disk} \propto r'_{\rm in}$$^{0}$).
Such variations in the hardening factor may 
in turn stem, e.g.,  from fluctuations of the disk height,
because this would affect the temperature gradient and/or
radiative transfer in the vertical direction of the disk.

In the VLB, the change in $r'_{\rm in}$ is considered real (\S~\ref{subsec:vari}),
because we observe a scaling of $kT'_{\rm in} \propto r'_{\rm in}$$^{-0.75}$
which agrees with the behavior of a standard accretion disk 
under a constant $\dot{M}$ and a constant hardening factor.
In this branch, 
the disk is considered to randomly retreat back from its ``home position",
which is located in Figure~\ref{fig:detrend} around $r'_{\rm in} \sim 0.85$
where the CLB and VLB meet.
When this takes place,
the accreting matter located between the retreated and the home-position radii 
will fall onto the NS surface without radiating the disk emission,
and the gravitational energy it acquires 
will be released solely as the radiation from the NS surface.
Then, $L_{\rm disk}$ will decrease while $L_{\rm BB}$ will increase.
This can naturally explain the anti correlation between the MCD and BB 
luminosities observed in the VLB (Figure~\ref{fig:detrend}).


\subsection{Conclusion}
The results of our study can be summarized in the following four points.
\begin{enumerate}
\item The physical picture provided by the Eastern model can 
consistently explain the behavior of 4U 1608--522 in the upper-banana state.
\item When the mass accretion rate increases, 
some fraction of the matter flowing through the accretion disk starts outflowing,
presumably due to an increased radiation pressure.
\item Even at the same accretion rate,
the spectral parameter changes randomly
on time scales of hours to days,
mainly reflecting changes in the innermost disk radius.
\item The change in the disk inner radius is a mixture of
real and apparent effects,
the former due to sporadic retreat of the disk
while the latter due to fluctuations in the color hardening factor.
\end{enumerate}

Facilities: \facility{RXTE (PCA)}.




\begin{table}[htbp]
\begin{center}
\caption
{Four spectra analyzed circumstantially in the upper-banana state.}
\label{tab:list}
\begin{tabular}{lcccc}
\hline 
Spectrum  & Observation & Count Rate & \multicolumn{2}{c}{Color} \\
          & ID          & [Crab]     & Soft$^{\rm a}$ & Hard$^{\rm b}$ \\
\hline 
Spec A & 30062-03-01-02 & 1.07 & 0.43 & 0.29 \\
\hline 
Spec B & 30062-03-01-01 & 1.03 & 0.41 & 0.26 \\
\hline 
Spec C & 30062-03-01-00 & 0.95 & 0.40 & 0.25 \\
\hline 
Spec D & 30062-03-01-04 & 0.78 & 0.42 & 0.27 \\
\hline 
\multicolumn{5}{@{}l@{}}{\hbox to 0pt{\parbox{160mm}{\footnotesize
$^{\rm a}$ The soft color refers to the source count ratio of 6--10 keV / 2.5--6 keV.\\
$^{\rm b}$ The same as the soft color but the ratio of 10--30 keV / 6--10 keV counts.\\
}\hss}}
\end{tabular}
\end{center}
\end{table}

\begin{table}[htbp]
\begin{center}
\caption
{Model fit results to the difference spectra calculated from those in Table~\ref{tab:list}.$^{\rm a}$}
\label{tab:spec_1608_upper}
\begin{tabular}{llcccc}
\hline 
Spectrum  & Model & $N_{\rm H}$~$^{\rm b}$ & $kT_{\rm BB}$~$^{\rm c}$ & $kT_{\rm in}$~$^{\rm c}$ & $\chi^2$/d.o.f.\ \\
\hline 
Spec 1 & BB  & $27^{+12}_{-11}$ & $2.5 \pm 0.1$ & $\cdots$ & 20/58 \\
(A - B) \\ 
\hline 
Spec 2 & BB  & $< 0.3$ & $1.27 \pm 0.02$ & $\cdots$ & 47/58 \\
(C - D) & MCD & $1.8 \pm 1.2$ & $\cdots$ & $1.7 \pm 0.1$ & 29/58 \\
\hline 
\multicolumn{6}{@{}l@{}}{\hbox to 0pt{\parbox{160mm}{\footnotesize
$^{\rm a}$ All the errors are single-parameter 90\% confidence limits.\\
$^{\rm b}$ The absorption column density in the unit of 10$^{22}$ cm$^{-2}$.\\
$^{\rm c}$ Temperatures are all in the unit of keV.\\
}\hss}}
\end{tabular}
\end{center}
\end{table}

\begin{table}[htbp]
\begin{center}\small
\caption
{Fitting results of the four original spectra in Table~\ref{tab:list} with the MCD+BB+Gau model.$^{\rm a}$}
\label{tab:spec_1608_upper2}
\begin{tabular}{lccccccc}
\hline 
Spectrum & \multicolumn{2}{c}{MCD} & \multicolumn{2}{c}{BB} & \multicolumn{2}{c}{Gau$^{\rm b}$} & \\
\cline{2-3}\cline{4-5}\cline{6-7}
 & $kT_{\rm in}$~$^{\rm c}$ & $r_{\rm in}$~$^{\rm d}$ & $kT_{\rm BB}$~$^{\rm c}$ & $r_{\rm BB}$~$^{\rm e}$ & Cen.\ Ene.$^{\rm f}$  & Norm$^{\rm g}$  &  $\chi^2$/d.o.f.\ \\
\hline 
Spec A  & $1.84^{+0.03}_{-0.04}$ & $4.1^{+0.1}_{-0.2}$ & $2.67 \pm 0.04$ & $1.5 \pm 0.1$ & $6.6 \pm 0.2$ & $1.9^{+0.5}_{-0.4}$ & 29/54 \\
\hline 
Spec B  & $1.88 \pm 0.04$ & $4.0^{+0.1}_{-0.2}$ & $2.78^{+0.1}_{-0.08}$ & $1.1 \pm 0.1$ & $6.6 \pm 0.2$ & $2.0 \pm 0.4$ & 41/54 \\
\hline 
Spec C  & $1.79 \pm 0.04$ & $4.2^{+0.2}_{-0.1}$ & $2.64 \pm 0.08$ & $1.2 \pm 0.1$ & $6.6^{+0.2}_{-0.1}$ & $1.7 \pm 0.4$ & 32/54 \\
\hline 
Spec D  & $1.84 \pm 0.06$ & $3.5^{+0.3}_{-0.2}$ & $2.7 \pm 0.1$ & $1.1^{+0.2}_{-0.1}$ & $6.6 \pm 0.2$ & $1.1^{+0.4}_{-0.3}$ & 42/54 \\
\hline 
\multicolumn{8}{@{}l@{}}{\hbox to 0pt{\parbox{160mm}{\footnotesize
$^{\rm a}$ All the errors are single-parameter 90\% confidence limits.
The absorption column density is fixed at $0.8 \times 10^{22}$ cm$^{-2}$.\\
$^{\rm b}$ The width of the Gaussian component is not well determined and constraint to be $\le 0.2$ keV.\\
$^{\rm c}$ Temperatures are all in the unit of keV.\\
$^{\rm d}$ The innermost radius in the unit of $\sqrt{\cos i}^{-1}$ km, where $i$ is the inclination angle.\\
$^{\rm e}$ The radius of the BB emission in the unit of km, assuming the emission comes from the isotropic sphere region.\\
$^{\rm f}$ The center energy of the Gaussian model, which is limited in the range of 6.4--6.9 keV.\\
$^{\rm g}$ The normalization of the Gaussian model in the unit of $10^{-2}$ photons cm$^{-2}$ s$^{-1}$.\\
}\hss}}
\end{tabular}
\end{center}
\end{table}

\begin{table}[htbp]
\begin{center}\small
\caption
{de-trended parameters of Spec A to D after eliminating the $L_{\rm tot}$ dependence in Figure~\ref{fig:detrend}.$^{\rm a,b}$}
\label{tab:detrend}
\begin{tabular}{lcccccc}
\hline 
 & $L'_{\rm disk}$ & $L'_{\rm BB}$ & $kT'_{\rm in}$ & $r'_{\rm in}$ & $kT'_{\rm BB}$ & $r'_{\rm BB}$ \\
\hline 
Spec A  & 1.69 & 1.11 & 1.43 & 0.97 & 1.58 & 1.15 \\
\hline 
Spec B  & 1.96 & 0.75 & 1.49 & 0.96 & 1.65 & 0.86 \\
\hline 
Spec C  & 1.93 & 0.80 & 1.44 & 1.04 & 1.58 & 0.97 \\
\hline 
Spec D  & 1.90 & 0.85 & 1.53 & 0.88 & 1.64 & 0.92 \\
\hline 
\multicolumn{7}{@{}l@{}}{\hbox to 0pt{\parbox{160mm}{\footnotesize
$^{\rm a}$ See equation~(\ref{eq:detrend}) for the $L_{\rm tot}$-dependence eliminated here.\\
$^{\rm b}$ Absolute values of the de-trended parameters are not meaningful.\\
}\hss}}
\end{tabular}
\end{center}
\end{table}

\newpage

\end{document}